\documentclass[twocolumn]{aastex631}

\graphicspath{{./}{figures/}}

\begin{document}

\title{The Mass of the Black Hole in NGC\,5273 from Stellar Dynamical Modeling}

\author[0000-0001-6771-1809]{Katie A. Merrell}
\affiliation{Department of Physics and Astronomy, Georgia State University, Atlanta, GA 30303, USA}

\author[0000-0002-5038-9267]{Eugene Vasiliev}
\affiliation{Institute of Astronomy, Madingley Road, Cambridge, CB3 0HA, UK}

\author[0000-0002-2816-5398]{Misty C. Bentz}
\affiliation{Department of Physics and Astronomy, Georgia State University, Atlanta, GA 30303, USA}

\author[0000-0002-6257-2341]{Monica Valluri}
\affiliation{Department of Astronomy, University of Michigan, Ann Arbor, MI 48104, USA}

\author[0000-0003-0017-349X]{Christopher A. Onken}
\affiliation{Research School of Astronomy and Astrophysics, Australian National University, Canberra ACT 2611, Australia}

\begin{abstract}
We present a new constraint on the mass of the black hole in the active S0 galaxy NGC\,5273. Due to the proximity of the galaxy at $16.6 \pm 2.1$\,Mpc, we were able to resolve and extract the bulk motions of stars near the central black hole using AO-assisted observations with Gemini NIFS, as well as constrain the large-scale kinematics using re-reduced archival SAURON spectroscopy. High resolution \textit{HST} imaging allowed us to generate a surface brightness decomposition, determine approximate mass-to-light ratios for the bulge and disk, and obtain an estimate for the disk inclination. We constructed an extensive library of dynamical models using the Schwarzschild orbit-superposition code FORSTAND, exploring a range of disk and bulge shapes, halo masses, etc. We determined a black hole mass of $M_{\bullet} = [0.5 - 2] \times 10^{7}$\,$M_{\odot}$, where the low side of the range is in agreement with the reverberation mapping measurement of $M_{\bullet} = [4.7 \pm 1.6] \times 10^{6}$\,$M_{\odot}$. NGC\,5273 is one of only a small number of nearby galaxies hosting broad-lined AGN, allowing crucial comparison of the black hole masses derived from different mass measurement techniques. 

\end{abstract}

\keywords{Stellar dynamics (1596) --- AGN host galaxies (2017) --- Seyfert galaxies (1447) --- Supermassive black holes (1663)}

\section{Introduction} \label{sec:intro}

The tight correlations that exist between supermassive black hole (SMBH) mass ($\sim10^{5}-10^{10}\,M_{\sun}$, see review by \citealt{bambi2018}) and host galaxy properties such as bulge stellar velocity dispersion (\citealt{ferrarese2000,gebhardt2000}), mass \citep{haring2004}, and luminosity \citep{marconi2003} imply that SMBHs are crucial components of galaxy evolution. Feedback effects from accretion onto the central black hole are observed to impact the host galaxy and are suspected to play a role in regulating galaxy growth (e.g., \citealt{silk1998}). Although the exact mechanisms involved in the co-evolution of these objects are not yet fully understood, accurate black hole mass measurements are fundamental pieces of the puzzle.  

The SMBH residing at the center of the Milky Way galaxy, Sagittarius A$^{\ast}$, maintains the most accurate mass measurement on record. Its nearby location allows precise monitoring of line-of-sight velocities and proper motions of individual stars within the potential well of the black hole (e.g., \citealt{ghez00,genzel00,schodel02}), with the most recent measurement of the black hole mass by the \citet{gravity2022} providing a value of $M_{\bullet} = [4.297 \pm 0.012] \times 10^{6}\,M_{\odot}$ (statistical uncertainty). With the current level of technology and instrumentation, this is the only SMBH with a proximity that allows some of the innermost individual stars to be resolved.

For all other galaxies, a different approach is needed, and several have been developed over the last few decades. Water maser emission from the thin circumnuclear gas disk orbiting only fractions of a parsec from the central black hole in NGC\,4258 is a powerful tracer of the Keplerian rotation curve (e.g., \citealt{miyoshi1995}). However, very few galaxies contain observable water masers \citep[e.g.,][]{panessa2020}, especially given that the disk must be viewed almost perfectly edge-on. Ionized (e.g., \citealt{macchetto1997}), warm (e.g., \citealt{hicks2008}), and cold \citep[e.g.,][]{davis2013,barth2016} gas dynamical modeling can be used to simulate the kinematic structure of a nuclear gas disk traced by spatially-resolved emission line measurements, from which a black hole mass can be constrained. The difficulty with hot or warm gas is that it must display circular rotation to be modeled accurately, however in practice non-gravitational perturbations, which may be large, are often seen (e.g., \citealt{verdoes2006}). Such gas is also affected by dust obscuration (e.g., \citealt{garcia2015}). Cold molecular gas generally exhibits less turbulence than warm or ionized gas, but it is unclear how widespread is the presence of cold, rotating nuclear gas disks, especially in the case of actively accreting SMBHs (e.g., \citealt{kakkad2017}). Currently, the most commonly used SMBH mass determination techniques are reverberation mapping \citep[e.g.,][]{blanford1982,cackett2021} and stellar dynamical modeling \citep[e.g.,][]{vandermarel1998,gebhardt2003,valluri2004,vandenbosch2010}. 

While stellar dynamical modeling is performed in the spatial domain, reverberation mapping utilizes measurements in the time domain. Reverberation mapping can only be used for galaxies with luminous broad-lined active galactic nuclei (AGN). The time delay between variations in emission from the continuum (which is interpreted as arising from the accretion disk) and the echo of those variations in the broad-line region (BLR) emission establishes the size of the BLR \citep{peterson1993}. By monitoring the continuum and broad-line variations, the derived radius and velocity of the gas clouds provide the elements to determine the enclosed black hole mass (\citealt{peterson1999,peterson2000}). This method does not require spatially resolving components of the galaxy and therefore does not depend on nor is limited by the distance to the galaxy.

At the present time, stellar dynamical modeling is restricted to galaxies within about 100\,Mpc \citep{gultekin2009} because it relies on the ability to spatially resolve the bulk motions of stars in the vicinity of the SMBH. The black hole mass also linearly depends on the assumed distance because of the need to convert angular scales on the sky to physical scales in the galaxy. Though individual stars cannot be resolved, spatially-resolved spectroscopy allows the bulk stellar kinematics at different spatial positions to be parameterized with the line-of-sight velocity distribution (LOSVD). High resolution photometry is also needed to accurately map the light distribution and establish the stellar contribution to the total gravitational potential traced by the kinematics. The black hole mass is determined from dynamical models built to simulate the observed surface brightness profile and observed LOSVDs \citep[see review by][]{kormendy2013}. 

The different inherent assumptions associated with various mass measurement techniques present a crucial reason why it is important to compare the mass of a black hole derived from multiple methods, as certain assumptions could cause systematic disagreements in the results. The giant elliptical galaxy M87 is a documented example of inconsistent constraints on the mass of the central black hole. The mass from geometric models and relativistic magnetohydrodynamic simulations of the emission ring near the event horizon ($M_{\bullet} = [6.5 \pm 0.7] \times 10^{9}$\,$M_{\odot}$; \citealt{eht2019}) is similar to the the stellar dynamical modeling mass ($M_{\bullet} = [6.6 \pm 0.4] \times 10^{9}$\,$M_{\odot}$; \citealt{gebhardt2011}), but conflicts with the gas dynamical modeling mass ($M_{\bullet} = [3.5^{+0.9}_{-0.7}] \times 10^{9}$\,$M_{\odot}$; \citealt{walsh2013}). This contradiction has been suspected to be due to the assumption that the thin nuclear gas disk exhibits circular Keplerian motion. Indeed, when investigating the effects of non-Keplerian orbits on black hole masses derived from gas dynamical modeling, \citet{jeter2019} were able to simulate velocity curves that closely matched the observations of M87 (including velocity dispersion) and would predict a larger black hole mass of $M_{\bullet} = 6.6 \times 10^{9}$\,$M_{\odot}$.

While reverberation mapping and stellar dynamical modeling are often used for measuring black hole masses, there have been very few cases where they have been applied to the same black holes. Studies of nuclear stellar dynamics have generally avoided AGN because they act as bright sources of noise at the location of the central black hole, and in the local universe, they are most often found in late-type galaxies, which makes them more difficult to model. Furthermore, there are only a handful of nearby galaxies with broad-lined AGN. Yet, direct comparisons of mass measurements from these two techniques are essential for investigating any potential biases in the black hole masses, especially given that the assumptions and observations employed by these two methods are completely independent of each other. 

To date, the only two galaxies that have been studied with both reverberation mapping and stellar dynamical modeling are NGC\,4151 (\citealt{bentz2006,onken2014,derosa2018,roberts2021,bentz2022}) at $D = 15.8 \pm 0.4$\,Mpc \citep{yuan2020} and NGC\,3227 (\citealt{davies2006,denney2010,derosa2018}) at $D = 23.7 \pm 2.6$\,Mpc (the distance to its interacting companion galaxy NGC\,3226; \citealt{tonry2001}).

NGC\,5273 is another nearby galaxy containing a broad-lined AGN, and thus provides an additional opportunity to compare reverberation and stellar dynamical black hole masses. A reverberation mass of $M_{\bullet} = [4.7 \pm 1.6] \times 10^{6}$\,$M_{\odot}$ has already been measured by \citet{bentz2014}. The radius of the black hole sphere of influence given by $r_{\rm infl}=GM_{\bullet}/\sigma_{\ast}^{2}$ was estimated to be about 3.7\,pc or $0\farcs05$ using the bulge velocity dispersion $\sigma_{R_{e}/8} = 74.1 \pm 3.7\,\mathrm{km\:s^{-1}}$ from \citet{cappellari2013b}. In this paper we present the nuclear stellar kinematics and a black hole mass derived from stellar dynamical modeling.

\section{Observations and Reductions}
 
\begin{figure}
\centering
\includegraphics{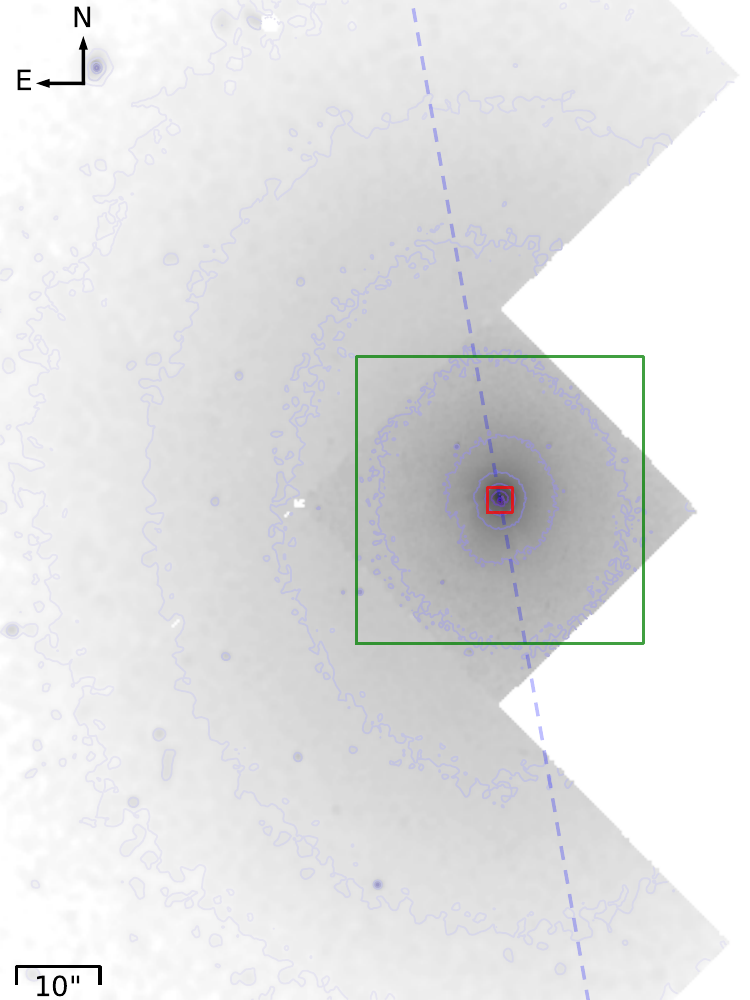}
\caption{Illustration of photometric and spectroscopic data coverage. The grayscale image shows the HST WFPC2 $F547M$ image rotated so that north is up, with contours spaced by 1 magnitude. Red and green rectangles show the NIFS and SAURON datasets respectively. The dashed blue line shows the orientation of the kinematic major axis.} 
\label{fig:illustration}
\end{figure}

NGC\,5273 is a Seyfert 1.5 \citep{trippe2010} in an unbarred SA0(s) galaxy \citep{devaucouleurs1991} with coordinates of $\alpha = 13^{\mathrm{h}}42^{\mathrm{m}}08^{\mathrm{s}}.3$ and $\delta = +35\degr39\arcmin15\arcsec$ \citep{clements1983}. An assessment of the galaxy surface brightness fluctuations provides a distance of $16.60 \pm 2.1$\,Mpc (\citealt{tonry2001} with the revisions of \citealt{tully2016}). 

\subsection{NIFS}

Observations were collected with the Gemini North Near-infrared Integral Field Spectrograph (NIFS; \citealt{mcgregor2003}) under program GN-2015A-Q-30 (PI: Bentz). The Laser Guide Star (LGS) Adaptive Optics (AO) system was utilized to improve spatial resolution by correcting real-time distortions due to atmospheric refraction. A total of 175 galaxy images, each with an exposure time of 120\,s, were gathered over seven nights between 8 July 2015 and 18 April 2016, resulting in about 5.8 hours of on-source time. The median airmass over all nights was 1.06 and no exposures were collected at airmasses larger than 1.56. The instrument was set to an on-sky position angle of $0^{\circ}$ and the $3\arcsec \times 3\arcsec$ field-of-view (FOV) was centered on the AGN (see Figure~\ref{fig:illustration}), covering approximately the inner $\rm 241\,pc \times 241\,pc$ in the galaxy nucleus. The NIFS image slicer divided the FOV into 29 vertical slices, each $0\farcs103$ across. The $3\arcsec$ slices were sampled length-wise by 69 detector pixels, resulting in a height of $0\farcs043$ for each pixel. The 2D image slices were stacked onto a Rockwell HAWAII-2RG HgCdTe detector. The $K$ grating and $HK$ filter yielded a spectral range of $1.99 - 2.40\,\mu$m and supplied a resolving power of 5290 and a velocity resolution of 56.7\,$\mathrm{km\:s^{-1}}$. From the argon-xenon arc lamp lines, we measured an instrumental resolution of 3.2\,{\AA}.

Variable absorption by Earth's atmosphere is especially prominent in the near-infrared, greatly affecting spectra observed at these wavelengths. Standard stars in close on-sky proximity to the galaxy were observed with the same instrument during each observing block. A-type stars are typically chosen due to their lack of metal lines, relative brightness, and continua that can be reasonably approximated by a blackbody \citep{vacca2003}. Therefore, observations of one or two A0V standard stars were captured every night to be used for telluric corrections. At the beginning of each observing block, four frames of HD109615 were obtained, each with an exposure time of 15\,s, and at the end of each observing block (except one), we obtained four frames of HD128039, each with an exposure time of 30\,s. The typical object-sky-object-object-sky-object dithering pattern for both galaxy and telluric observations permitted sky images to be subtracted from chronologically adjacent object images.

A summary of the galaxy and standard star observations is given in Table \ref{table1}. The argon-xenon arclamp images used for wavelength calibrations were collected at the end of each block of galaxy observations. The Ronchi flats used for spatial calibrations, darks, and flats were typically acquired as part of the end-of-night calibrations.

We employed a new Gemini NIFS reduction pipeline\footnote{developed by Jonelle Walsh, Anil Seth, Richard McDermid, Nora Luetzgendorf, and Mariya Lyubenova}$^{,}$\footnote{https://github.com/remcovandenbosch/NIFS-pipeline}, which included sky subtraction, flat fielding, dark subtraction, bad pixel masking, wavelength and spatial calibration, and telluric corrections. Individual frames were corrected to heliocentric velocities, rectified with the distortion/dispersion solution, and aligned/combined into a data cube (and noise cube) with a re-sampled pixel scale of $0\farcs05 \times 0\farcs05$ and constant 2.13\,{\AA}$\mathrm{\:pix^{-1}}$ spectral dispersion. The new pipeline contained capabilities not yet implemented in the widely-used Gemini NIFS IRAF reduction scripts\footnote{www.gemini.edu/instrumentation/nifs/data-reduction} such as variance propagation, heliocentric velocity corrections, and aligning/combining frames. Additional developments included improving several NIFS IRAF tasks, revising the method for removing telluric absorption, and increasing automation. A comparison of the results from the new pipeline and the original Gemini NIFS IRAF reduction scripts was carried out by \citet{merrell2020}. We made a few modifications to the new pipeline in an effort to reduce some of the unmitigated noise seen in the data. First, we measured the full width at half maximum (FWHM) of the galaxy nucleus in each calibrated frame to identify any observations where the AO correction was poor. We discarded 15 frames with FWHM values larger than 5.5 pixels, compared to the median FWHM of 4.1 pixels among the remaining frames. During the frame combining step, we added a second wave of bad pixel masking with the IRAF task \texttt{crmedian}. Pixels with values that exceeded five times the standard deviation of a nearby block of pixels were flagged and masked. Finally, we median scaled each frame to match the first frame of the first night. Our final data cube is the combination of 160 individual galaxy frames and has an effective total exposure time of 5.33 hours. 

A point-spread function (PSF) image was created from the final data cube by combining several consecutive 2D image slices at wavelengths corresponding to the strong broad Br\,$\gamma$ AGN emission line and subtracting a slice that was representative of the continuum underlying the line emission. The resulting image of the spatially unresolved broad-line region emission thus represents the PSF of the final combined galaxy cube. The PSF image was well-described by two circular Gaussians with common centers and nearly equal weight. The best-fit pair of Gaussians includes a compact Gaussian representing the diffraction-limited AO core and a wider Gaussian representing the uncorrected seeing halo. Table \ref{table2} lists the Gaussian widths and flux weights of the two Gaussians that we determined using GALFIT (\citealt{peng2002,peng2010}).

\begin{deluxetable}{lcccc}
\caption{NIFS Observations
\label{table1}}
\tablehead{
\colhead{UT Date} & \colhead{Target} & \colhead{Exposure Time} & \colhead{$\#$ of Frames} \\
\nocolhead{} & \nocolhead{} & \colhead{(s)} & \nocolhead{}
}
\startdata
 2015 Jul 8 & HD 109615 & 15 & 4 \\
 & NGC\,5273 & 120 & 32 \\
 & HD 128039 & 30 & 4 \\
 2016 Feb 24 & HD 109615 & 15 & 4 \\
 & NGC\,5273 & 120 & 16 \\
 2016 Feb 25 & HD 109615 & 15 & 4 \\
 & NGC\,5273 & 120 & 33 \\
 & HD 128039 & 30 & 4 \\
 2016 Feb 27 & HD 109615 & 15 & 4 \\
 & NGC\,5273 & 120 & 32 \\
 & HD 128039 & 30 & 4 \\
 2016 Feb 29 & HD 109615 & 15 & 4 \\
 & NGC\,5273 & 120 & 23 \\
 & HD 128039 & 30 & 4 \\
 2016 Mar 4 & HD 109615 & 15 & 4 \\
 & NGC\,5273 & 120 & 16 \\
 & HD 128039 & 30 & 4 \\
 2016 Apr 18 & HD 109615 & 15 & 4 \\
 & NGC\,5273 & 120 & 23 \\
 & HD 128039 & 30 & 4 \\
\enddata
\end{deluxetable}

\begin{deluxetable}{lccc}
\caption{PSF\label{table2}}
\tablehead{
\colhead{$\sigma$ (arcsec)} & \nocolhead{} & \colhead{Weight}
}
\startdata
& NIFS & \\
\hline
0.076 & & 0.53 \\
0.29 & & 0.47 \\
\hline
& SAURON & \\
\hline
0.32 & & 0.50 \\
0.76 & & 0.50 \\
\enddata
\end{deluxetable}

\subsection{SAURON}

The wider-field William Herschel Telescope (WHT) Spectroscopic Areal Unit for Research and Optical Nebulae (SAURON; \citealt{bacon2001}) integral field spectrograph observations of NGC\,5273 were collected as part of the volume-limited ATLAS$^{\textrm{3D}}$ project that examined stellar and gas kinematics and photometric imaging of 260 early-type galaxies \citep{cappellari2011}. The SAURON $33\arcsec \times 41\arcsec$ FOV was sampled by $0\farcs94$ square lenslets in the low resolution (LR) mode. The instrument major axis was aligned with the photometric major axis of the galaxy ($8.9^{\circ} \pm 1.0^{\circ}$ East of North; \citealt{krajnovic2011}), but the datacube provided at the ATLAS$^{\rm 3D}$ website\footnote{https://www-astro.physics.ox.ac.uk/atlas3d/} is rebinned onto a grid aligned with the north direction and re-sampled to pixel size $0\farcs8 \times 0\farcs8$. The spectral coverage of $4800 - 5400$\,{\AA} was chosen so that the H$\beta$ and [\ion{O}{3}] emission lines could be studied and the Mg $b$ absorption lines could be used for stellar kinematic analysis \citep{bacon2001}. The LR mode provided a spectral resolution of $\rm FWHM = 4.2$\,{\AA} and an instrument dispersion of 105\,$\mathrm{km\:s^{-1}}$. Observations for the ATLAS$^{\textrm{3D}}$ collaboration were gathered over 38 nights between 10 April 2007 and 11 March 2008, with two on-source hours dedicated to each galaxy in the sample \citep{cappellari2011}. Image reductions were performed by the ATLAS$^{\textrm{3D}}$ team using the \texttt{XSAURON} software \citep{bacon2001}. The reduction process included bias and dark subtraction, flat fielding, removal of cosmic rays, wavelength calibration, sky subtraction, and flux calibration. 

We downloaded the final reduced and combined image cube for NGC5273 from the ATLAS$^{\textrm{3D}}$ website. The final FOV was cropped to $33\arcsec \times 33\arcsec$ (see Figure \ref{fig:illustration}) in order to fully symmetrize the kinematics. To characterize the SAURON PSF, we followed a process similar to the one described by \citet{emsellem2004}. We compared an image of NGC\,5273 obtained with the \textit{Hubble Space Telescope (HST)} Wide-Field and Planetary Camera 2 (WFPC2) $F547M$ filter to a 2D spatial slice extracted from the SAURON data cube at a wavelength corresponding to continuum emission. The \textit{HST} image was first rotated, rebinned, and scaled in intensity to match the SAURON image. We then created and explored a grid of kernels made from two round Gaussians with common centers and weights, but different widths. Each kernel was used to blur the \textit{HST} image, after which it was subtracted from the SAURON image and the magnitude of the residuals was recorded. The kernel that was associated with the smallest residuals between the SAURON image and the blurred \textit{HST} image was determined to provide the best representation of the SAURON PSF, and we list the widths of the Gaussians for this kernel in Table \ref{table2}.

\section{Kinematics}

We extracted the stellar kinematics from the NIFS and SAURON data cubes using the Penalized Pixel-Fitting (pPXF) software (\citealt{cappellari2004,cappellari2017}) which parametrically recovers the LOSVD from an observed galaxy spectrum directly in pixel space. For each observed spectrum, a model galaxy spectrum is generated using a linear combination of stellar template spectra. pPXF then adjusts the LOSVD with which the model spectrum is convolved, and compares the convolved model with the observed spectrum. A best-fit LOSVD is determined by minimizing the $\chi^{2}$ between the observed and modeled galaxy spectrum, and for a data cube, the process is simply looped over all of the individual spectra collected at each spatial position.

Although the LOSVD is generally well-represented by a Gaussian function, pPXF utilizes a Gauss--Hermite series to represent the LOSVD shape in more detail when the signal-to-noise (S/N) is high. pPXF sets $h_1=h_2=0$ and determines best-fitting mean and width of the Gaussian profile $v,\sigma$ and the higher-order moments $h_3$--$h_6$. The $h_{3}$ and $h_{5}$ moments depict the skewness and other asymmetric deviations from a Gaussian profile, and the $h_{4}$ and $h_{6}$ moments depict the kurtosis and other symmetric deviations. As the name suggests, pPXF penalizes the LOSVD fit toward a more Gaussian shape when the S/N is low and the data do not contain sufficient information to be parameterized by the higher order Gauss--Hermite terms. pPXF allows simultaneous fitting of spectra from symmetric spatial positions instead of fitting each spectrum separately, and improvements in the fits can be made by including additive Legendre polynomials to help fix template mismatch and sky subtraction errors, and multiplicative Legendre polynomials to help fix reddening and flux calibration errors \citep{cappellari2017}.

The Voronoi binning routine of \citet{cappellari2003} was used to optimize the S/N across the FOV and limit any biases in the fits produced by regions of low S/N. The Voronoi tesselation method bins adjacent spatial pixels (spaxels), maintaining approximately constant S/N per bin across the entire image. The innermost regions usually retain one spaxel per bin, whereas the bins become larger further out as the surface brightness decreases.

\subsection{Symmetrization}

The velocity field of an axisymmetric dynamical model has fourfold discrete symmetry (reflection about the major axis leaves the velocity unchanged, and reflection about the minor axis flips its sign), thus is fully specified by one quadrant of the entire image.
If the provided observational data are not symmetrized, the model will effectively try to ``find the middle ground'' between the measurements in different quadrants, so in principle it is not necessary to symmetrize the data beforehand. However, there are two reasons why this symmetrization is beneficial. First, the pPXF routine can simultaneously fit two spectra in mirror-symmetric bins with a single LOSVD (flipped in sign in one of the two bins), and this produces more reliable results than fitting both spectra separately and averaging the results, because of increased S/N in the joint fit. Naturally, the uncertainty in the joint fit will be on average lower by a factor $\sim\sqrt{2}$. Second, when the input kinematic maps cover only one or two quadrants instead of four, correspondingly reducing the number of independent measurements and their uncertainties, the orbit-superposition modelling is more efficient, since fewer constraints need to be satisfied.

In our case, since the footprints of both NIFS and SAURON data are not aligned with the kinematic major axis of the galaxy, individual pixels in one quadrant do not line up with pixels in the adjacent quadrant (flipped about either of the two principal axes), thus it is not possible to combine the spectra in adjacent quadrants. However, the central pixel is placed exactly at the galaxy center, so when flipping about both axes simultaneously (equivalently, applying a mirror symmetry w.r.t.\ the origin), opposite pixels do line up. When running the Voronoi binning code, we restrict it to only one half-plane, and place identical bins symmetrically in the other half, thus making it possible to run the joint pPXF fits in mirror-symmetric bins (except the central pixel, which does not have a symmetric counterpart). By doing so, we still end up with two independent quadrants (reflected about the major axis), which are usually fed into the model simultaneously, but allow us to test the systematic variations by running models with only one of the two quadrants.

To construct the Voronoi binning, we first determine the kinematic position angle (KPA; \citealt{krajnovic2006}) which defines the line of maximum rotation across the velocity map. The same KPA was used for both the NIFS and SAURON binning patterns and was set to $9.4 \pm 2.1$ after comparing the KPA measured from the final NIFS data cube ($6.2^{\circ} \pm 12.4^{\circ}$), the KPA measured from the final SAURON data cube ($12.4^{\circ} \pm 3.1^{\circ}$), and the value reported by ATLAS$^{\textrm{3D}}$ ($190.5^{\circ} \pm 7.0^{\circ}$, equivalent to $10.5^{\circ} \pm 7.0^{\circ}$ for our purposes). For the SAURON kinematics, we re-fit the KPA using a different binning scheme than was used by the ATLAS$^{\textrm{3D}}$ collaboration.

\subsection{NIFS}

\begin{figure*}
\centering
\includegraphics{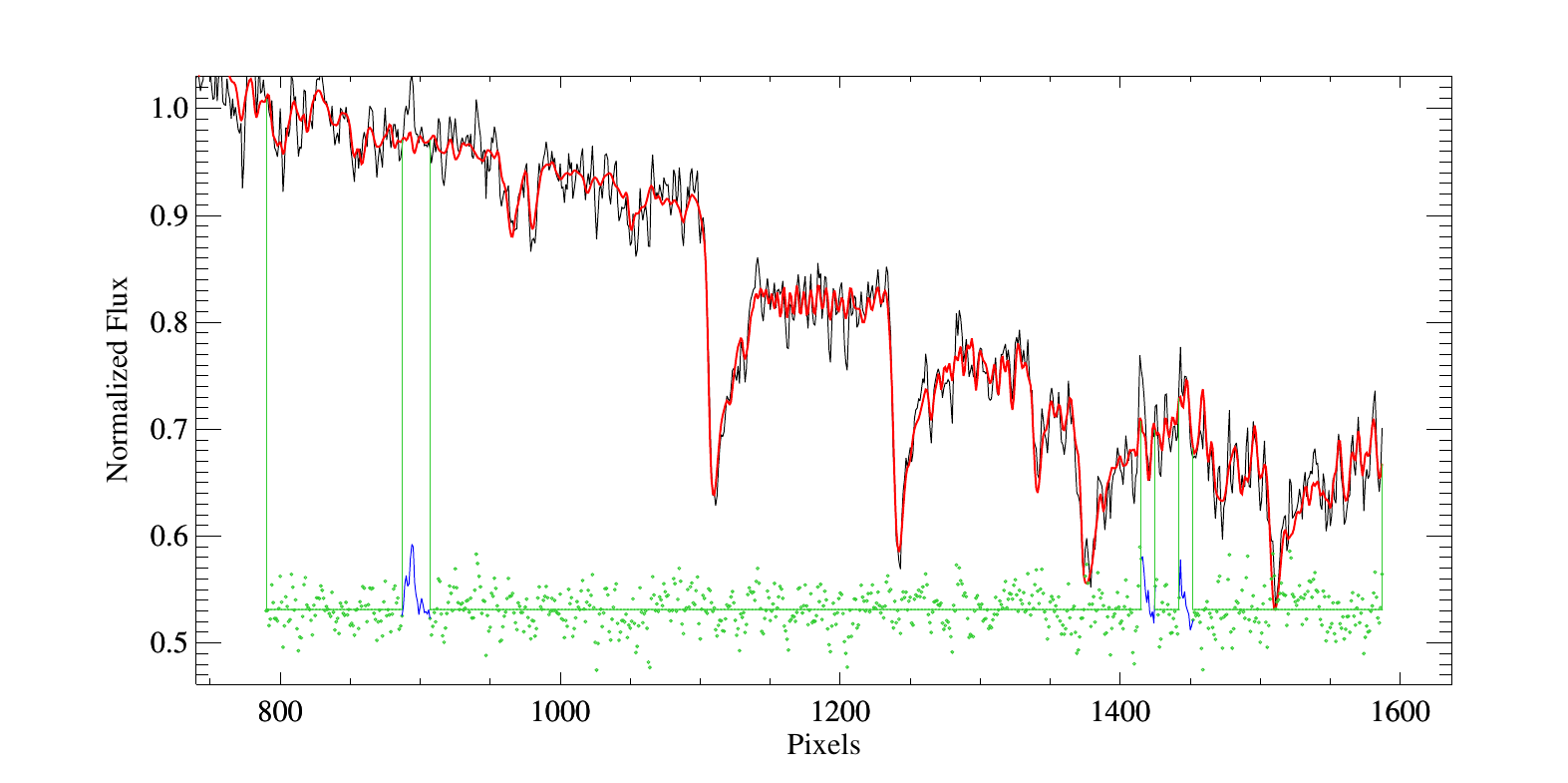}
\caption{Example of co-added NIFS spectra within a single bin in black with the pPXF fit overlaid in red, residuals (data - model) in green, and masked features in blue. The wavelengths span the range $2.24 - 2.41\,\mu$m. The pixels on the x-axis correspond to the array indices of the log rebinned wavelengths and the y-axis shows the normalized flux. The spatial location of the binned spectrum is about 0\farcs8 North of the galaxy center.}
\label{fig:spectrum}
\end{figure*}

The stellar templates used in the pPXF analysis of the NIFS observations of NGC\,5273 included 30 G, K, and M star spectra from the NIFS V1.5 and V2.0 template libraries\footnote{www.gemini.edu/observing/resources/near-ir-resources/spectroscopy/spectral-templates-library-v20} \citep{winge2009}. We restricted the spectral fits to a wavelength range of $2.24 - 2.41\,\mu$m to avoid the broad Br\,$\gamma$ AGN emission line ($2.17\,\mu$m) and to focus the fits on several strong CO band heads ($2.29 - 2.40\,\mu$m). We also masked out faint emission lines at the observed wavelengths 2.26\,$\mu$m, 2.37\,$\mu$m, and 2.38\,$\mu$m. An observed spectrum and accompanying pPXF fit belonging to a bin near the edge of the FOV can be seen in Figure \ref{fig:spectrum}. To more accurately determine the LOSVD variations as a function of spatial position, an optimal template solution for a single spaxel was identified from the best template solutions across all spaxels, compiled based on the lowest $\chi^{2}$ values associated with the spectrum fits. This optimal template was used to fit all the spatial bins in the NIFS data, implying a constant stellar population across the FOV (loosening this assumption would likely cause some kinematic changes). The optimal template was mainly comprised of a weighted combination of spectra from six stars with spectral classifications K0III, K0IIIb, K5III, K5Ib, M0III, and M5III.

\begin{figure*}
\centering
\includegraphics{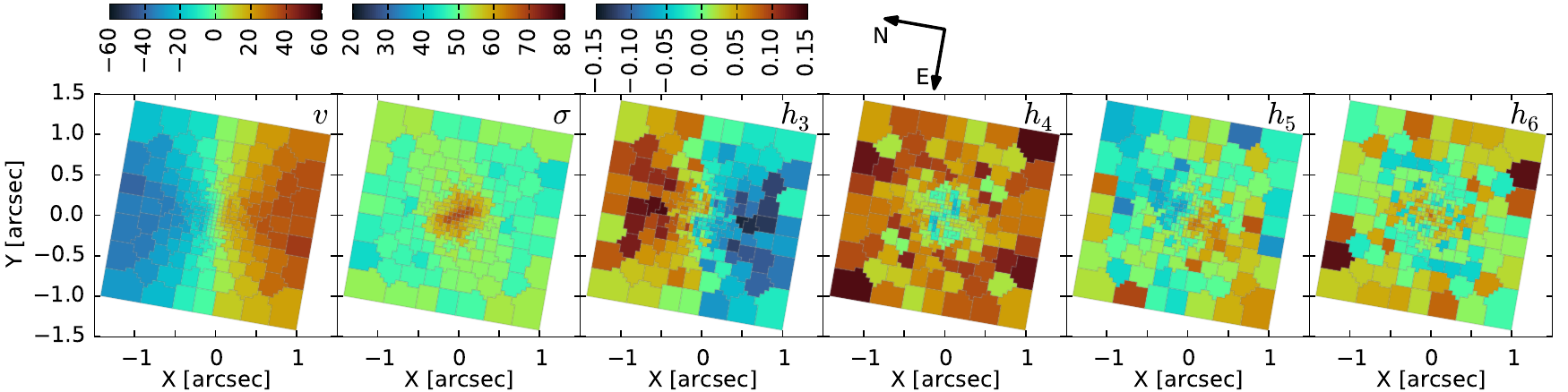}
\caption{Stellar kinematic maps derived from pPXF fits to the NIFS observations of NGC\,5273: central value $v$ and width $\sigma$ of the base Gaussian and coefficients of Gauss--Hermite expansion $h_3$--$h_6$. The maps are rotated so that the kinematic major axis is horizontal, and the orientation of north and east is indicated by arrows; the field of view is $2\farcs4 \times 2\farcs4$. The $v$ and $\sigma$ panels each have their own color bars and the Gauss--Hermite terms $h_3$--$h_6$ use the same color scaling. The velocity map shows a clear gradient indicative of rotation, and the $\sigma$ map has a clear peak in the center revealing the presence of the SMBH.}
\label{fig:nifs_maps}
\end{figure*}

The point-symmetric fits of the 193 bins were improved by adopting additive 4th order and multiplicative 2nd order Legendre polynomials. With the penalty term set to 0.2, the overall structure in each higher order Gauss--Hermite moment map was maintained, but allowed some smoothing near the edges. The $\chi^{2}$ values of the fits suggested that the noise spectra were slightly underestimating the uncertainties in the observations, so we increased the errors by a factor of 1.2 such that the median $\chi^{2}$ for the pPXF fits was $\sim 1.0$. The final NIFS kinematic maps are shown in Figure \ref{fig:nifs_maps}.

\subsection{SAURON}
 
Although kinematic results were provided by the ATLAS$^{\textrm{3D}}$ project, their Voronoi binning was not mirror-symmetrized, so we constructed our own symmetric binning scheme and reran the spectral fitting code on the new bins. Several libraries of optical stellar spectra exist, but none that were collected with SAURON. Instead, we started with the MILES template library of \citet{sanchezblazquez2006} with the updates and corrections of \citet{falconbarroso2011}. From the 985 template spectra, we selected a subset of 148 stars based on their spectral types (F0 to K7) and commonality with the Indo-US \citep{valdes2004} and ELODIE \citep{moultaka2004} libraries. The template spectra cover the wavelength range $3525 - 7500$\,{\AA} and have a well-determined spectral resolution of 2.51\,{\AA} \citep{beifiori2011}. 
 
The pPXF analysis of the SAURON observations was carried out in the optical band and includes the Mg $b$ absorption features between 5167\,{\AA} and 5184\,{\AA} in the rest frame. The SAURON spectra were fit covering a wavelength range of $4800 - 5400$\,{\AA} with the strong H$\beta$ $\lambda 4861$\,{\AA} and [\ion{O}{3}] $\lambda \lambda 4959,5007$\,{\AA} emission lines masked during the fitting process. The template resolution was degraded to match the instrumental resolution of 4.2\,{\AA} for the galaxy spectra \citep{cappellari2011}.
 
\begin{figure}
\centering
\includegraphics{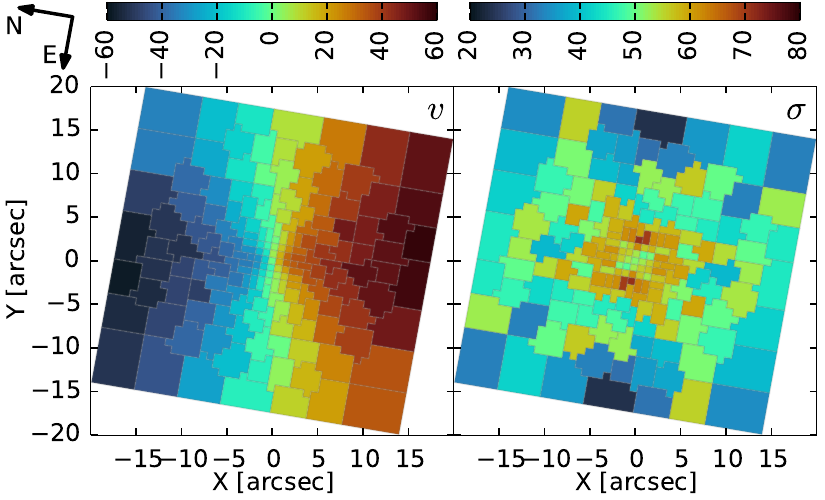}
\caption{Stellar kinematic maps derived from pPXF fits to the SAURON observations of NGC\,5273. The orientation and color scale are the same as in Figure~\ref{fig:nifs_maps}, but only $v$ and $\sigma$ were extracted from the spectra, setting higher-order Gauss--Hermite moments to zero.
}
\label{fig:sauron_maps}
\end{figure}

We fit only the velocity and velocity dispersion with pPXF because below $\sigma = 120\,\mathrm{km\:s^{-1}}$ (about 2 pixels), the higher order Gauss--Hermite terms could not be sufficiently constrained by the spectra observed with SAURON \citep{emsellem2004}, so with measured velocity dispersions for NGC\,5273 generally between $40 - 70\,\mathrm{km\:s^{-1}}$, only $V$ and $\sigma$ could be recovered reliably. Therefore, a penalty term was not set as it only affects the higher order Gauss--Hermite moments. Nevertheless, in the models we still constrained all six Gauss--Hermite moments, setting the dummy values for $h_3$--$h_6$ to zero with relatively large uncertainties of 0.1; this prevents the models from producing unrealistically wiggly LOSVDs. As with the NIFS data, we adopted a single optimal template to be used in fitting all of the SAURON spectra. The optimal template was mainly comprised of six stars with spectral classifications K0V, K0.5III, K1III, K3III, K3V, and K5III. The continuum fits were greatly improved by including 4th order additive and multiplicative Legendre polynomials in the point-symmetrized fitting routine for the final 181 bins. Finally, we inflated the uncertainties on the spectra by a factor of 1.6 to account for a slight underestimation and reduce the median $\chi^{2}$ to $\sim 1.0$. The point-symmetric SAURON kinematic maps are shown in Figure \ref{fig:sauron_maps}.

\section{Photometry}  \label{sec:photometry}

\begin{figure}
\includegraphics{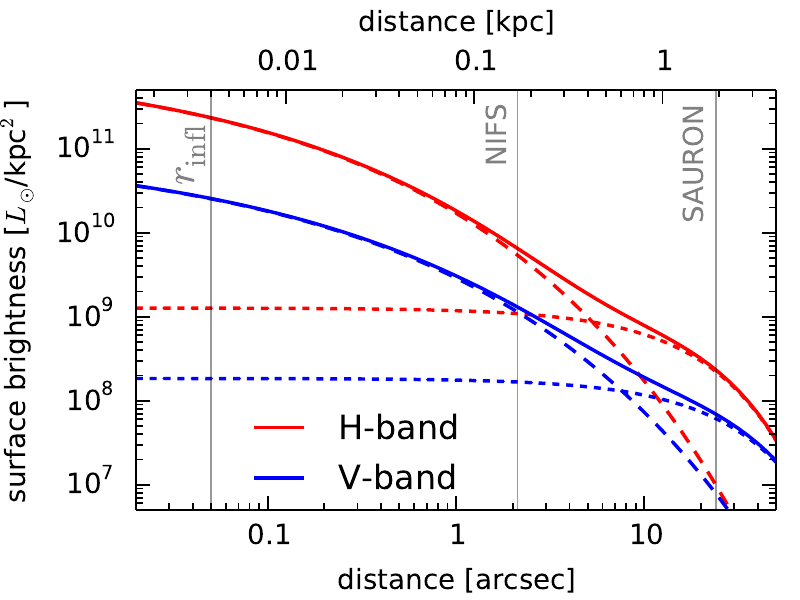}
\caption{One-dimensional surface brightness profiles derived from the $F547M$ HST image (blue) and the H-band image from NICFPS (red). Long- and short-dashed lines show the contributions from the bulge and the disk components, respectively, and solid lines show their sum. Vertical lines mark the radius of influence for a $5\times10^6\,M_\odot$ SMBH and the extent of the NIFS and SAURON kinematic datasets.}
\label{fig:density_profile}
\end{figure}

Not only does the dynamical modeling depend on the stellar kinematics, but it also depends on the stellar light distribution across the FOV. The total gravitational potential is affected by visible objects like orbiting stars as well as invisible objects such as the central black hole and dark matter halo. High resolution imaging is typically used to determine the surface brightness profile of stars in the galaxy nucleus. To create the 3D stellar mass density profile, we first fit the 2D surface brightness distribution with a sum of analytic profiles, then deproject the profiles associated with the host galaxy for a particular inclination angle to obtain 3D luminosity density profiles.  Finally, we multiply the 3D luminosity density profiles by the stellar mass-to-light ratio ($M/L$ or $\Upsilon$). These steps are described in more detail below.

\subsection{Surface Brightness Decomposition}  \label{sec:photometric_model}

We use GALFIT to construct analytic multi-component photometric models from the \textit{HST} image of NGC\,5273 through the $F547M$ filter. The surface brightness is characterized by a S{\'e}rsic profile for the bulge and an exponential profile for the disk, as well as a PSF image for the AGN and a gradient in $x$ and $y$ for the background sky. The parameterization of the host galaxy shown in Table \ref{table3} is defined by the $V$-band integrated magnitude ($m_{V}$), effective radius ($R_{e}$), S{\'e}rsic index ($n$), axis ratio ($q$), and position angle (PA) of each component. The bulge and disk components have common centers and position angles, with the bulge being more circular than the disk. It should be noted that although we refer to the inner component as ``the bulge'', it appears to be quite strongly flattened and rapidly rotating.  We adopted a VegaMag zeropoint of 21.651\,mag and applied a small color correction of $V - F547M = 0.0174$\,mag to convert the $F547M$ magnitudes to the $V$ band. Finally, we corrected the photometry for a Galactic extinction of $A_{V} = 0.028$\,mag using the \citet{schlafly2011} recalibration of the \citet{schlegel1998} Milky Way dust map. 

\begin{deluxetable}{lcccccc}
\caption{GALFIT Parameterization\label{table3}}
\tablehead{
\colhead{Comp.} & \colhead{$m_{V}$} & \colhead{$R_{e}$} & \colhead{$n$} & \colhead{$q$} & \colhead{PA}\\
\nocolhead{} & \colhead{(mag)} & \colhead{(\arcsec)} & \nocolhead{} & \nocolhead{} & \colhead{($^{\circ}$)}
}
\startdata
Bulge & 13.3 & 6.3 & 3.3 & 0.89 & 3.5\\
Disk & 11.8 & 40.2 & 1.0 & 0.83 & 3.5\\
\enddata
\tablecomments{Magnitudes are $V$-band using the VegaMag zeropoint and are corrected for Galactic extinction. At the distance of NGC\,5273, $1\arcsec$ corresponds to $\sim 80$\,pc. The PA is measured East of North.}
\end{deluxetable}

\subsection{Mass-to-Light Ratio}  \label{sec:photometric_ML}

The colors of stellar populations in galaxies are often used to predict stellar $M/L$ \citep{bell2001}. We therefore investigated the color as a function of radius from the galaxy center. The $F547M$ image (adjusted to $V$-band) was rotated, scaled, and blurred to match the spatial resolution and scale of an $H$-band image from the Apache Point Observatory Near-Infrared Camera \& Fabry-Perot Spectrometer. Elliptical isophotes were fit to each image, from which 1D surface brightness profiles were obtained. We determined $V-H = 3.1 \pm 0.1$\,mag for the bulge and $V-H = 2.6 \pm 0.1$\,mag for the disk. The bulge and disk colors were estimated from regions spanning $1\arcsec-3\arcsec$ and $10\arcsec-30\arcsec$ in radius, respectively. We calculated $\Upsilon_{V, V-H} = 5.2\pm0.9$ for the bulge and $\Upsilon_{V, V-H} = 2.1\pm0.4$ for the disk and therefore expect that the ratio $\mathcal R\equiv \Upsilon_{\rm disk}/\Upsilon_{\rm bulge}$ may be as low as 0.4. Figure~\ref{fig:density_profile} compares the azimuthally-averaged one-dimensional surface brightness profiles for both $V$ and $H$ bands: the latter is steeper in the central region, corresponding to a redder color and higher expected $M/L$.

We also estimated the $M/L$ values from several reported measurements of the color of NGC\,5273 using the equations derived from galaxy evolution models in \citet{bell2001}. We gathered $B-V$, $B-R$, and $V-K$ colors averaged for the whole galaxy from \citet{barway2005} and calculated $V$-band $M/L$ ratios of $\Upsilon_{V, B-V} = 2.5\pm0.2$, $\Upsilon_{V, B-R} = 2.3\pm0.1$, and $\Upsilon_{V, V-K} = 4.0\pm0.6$. From \citet{blakeslee2001}, we obtained $V-I$ color measured roughly at the half-light radius and calculated $\Upsilon_{V, V-I} = 3.3\pm0.2$. Using an average $B-V$ color assumed for the bulge and disk from \citet{prugniel1998} and a typical uncertainty of 0.1\,mag, we found $\Upsilon_{V, B-V} = 3.2\pm1.0$. These are consistent with the calculations for more defined regions of the bulge and disk. We therefore expect values of stellar $M/L$ between approximately 2 and 4 $M_{\odot}/L_{\odot}$. 

In comparison, \citet{cappellari2013a} determined $\Upsilon=3.3$ for the $r$ band, rather than $V$, within the effective half-light radius of the galaxy, $R_{e} = 38\farcs0 \pm 2\farcs3$, from JAM models.
This generally agrees with our estimates based on the galaxy color, even though $M/L$ is expected to differ as a function of the filter bandpass and it is not clear that the central AGN was accounted for in their analysis.

\subsection{Galaxy Inclination}  \label{sec:inclination}

In general, the problem of deprojection of a given 2D surface brightness profile into a 3D luminosity density profile does not have a unique solution even within the class of axisymmetric models \citep{gerhard1996,kochanek1996}. On the other hand, under the assumption that the 3D density is stratified on concentric ellipsoids with fixed axis ratios, the projected density is also ellipsoidally stratified \citep{contopoulos1956}, and the ellipsoidal 3D density can be uniquely reconstructed for any assumed orientation (within limits dictated by the physical validity of the solution), which explains the overwhelming prevalence of this assumption in the literature. In particular, for an oblate axisymmetric 3D density profile with the short-to-long axis ratio $Q\le 1$, the axis ratio of the projected ellipse is
\begin{equation}  \label{eq:projected_shape}
q = \sqrt{Q^2\,\sin^2i + \cos^2i},
\end{equation}
where $i$ is the inclination angle. Clearly, the lowest possible inclination $i_\mathrm{min} = \arccos q$ is given by projecting an infinitely thin disk, $Q=0$, and for any value $i_\mathrm{min} \le i \le 90^\circ$ the above equation can be inverted to determine $Q$ from $q$ and $i$. Because of a strong dependence of $Q$ on $i$ near the minimum value, we prefer to use the intrinsic disk axis ratio $Q$ as the independent parameter and determine $i$ from 
\begin{equation}  \label{eq:inclination}
\sin i =  \sqrt{\frac{1-q^2}{1-Q^2}}.
\end{equation}
In the case of NGC\,5273, the outermost component (the disk) has a projected axis ratio $q_{\rm disk}=0.83$, setting the lower limit on the inclination at $33.9^\circ$. 
We consider the intrinsic axis ratios $Q_{\rm disk}$ in the range 0.1--0.4 to be typical of disk galaxies (see e.g. Fig.1 in \citealt{sandage1970} or Fig.4 in \citealt{weijmans2014}), corresponding to $i=34.1^\circ-37.5^\circ$. 

The innermost bulge component is expected to share the same inclination angle as the disk, but since it is rounder in projection ($q_{\rm bulge} \simeq 0.89$), its 3D axis ratio must be larger (i.e., less flattened) than $Q_{\rm disk}$. In principle, it is uniquely determined from $q_{\rm bulge}$ and $i$ (the latter, in turn, computed from the observed $q_{\rm disk}$ and assumed $Q_{\rm disk}$), producing values of $Q_{\rm bulge}$ between 0.58 and 0.66 as $i$ varies between $34.1^\circ$ and $37.5^\circ$.
However, we find it beneficial to explore a wider range of bulge shapes $Q_{\rm bulge}$ independently of the disk shape $Q_{\rm disk}$ (equivalently, $i$). Effectively, this means that we do not use the exact value of $q_{\rm bulge}$ as determined by the best-fit GALFIT model, but take the liberty to consider values of $Q_{\rm bulge}$ in the range 0.4--0.7, which produce projected shapes $q_{\rm bulge}=0.83$--0.90 and allow for a few percent uncertainty on the measurement of $q_{\rm bulge}$. The motivation is that the dynamical models can be quite sensitive to the parameters of the density profile (including axis ratios), and we wish to consider how the uncertainties in the latter may affect the inference on $M_{\bullet}$. Moreover, the above derived relation between the intrinsic and projected axis ratios is strictly valid only for ellipsoidal profiles, and variations in $q_{\rm bulge}$ at a level of a few per cent are not implausible for more complex intrinsic shapes.

The above procedure is originally designed for ellipsoidally stratified profiles, however, galactic disk densities are often described by a different functional form separable in cylindrical coordinates:
\begin{equation}  \label{eq:disk_radial_profile}
\rho_{\rm disk}(R,z) = \frac{M}{2\pi\,R_{\rm disk}^2}\exp(-R/R_{\rm disk})\, h(z),
\end{equation}
with two alternative choices for the vertical profile:
\begin{equation}  \label{eq:disk_vertical_profile}
\begin{array}{ll}
\mbox{exponential} & h(z) = \frac{1}{2h_{\rm disk}} \exp\big(-|z|/h_{\rm disk}\big), \\
\mbox{isothermal} & h(z) = \frac{1}{4h_{\rm disk}} {\rm sech}^2\big(z/[2h_{\rm disk}]\big).
\end{array}
\end{equation}
In this case, the projected density is no longer exactly ellipsoidal, but can be still approximated as such, if we substitute $Q_{\rm disk} = 2h_{\rm disk} / R_{\rm disk}$ for the exponential or $Q_{\rm disk} = 2.5h_{\rm disk} / R_{\rm disk}$ for the sech$^2$ profile.

As part of the ATLAS$^{\textrm{3D}}$ project, inclinations were estimated from Jeans Anisotropic Multi-Gaussian Expansion (MGE) models (\citealt{emsellem2004}; \citealt{cappellari2008}), which revealed an inclination of $35^{\circ}$ for NGC\,5273 measured at the effective half-light radius \citep{cappellari2013a}. \citet{gutierrez2011} determined the apparent axis ratio $q_{\rm disk}$ of the outer disk by fitting ellipses to galaxy isophotes derived from images obtained with the Wide Field Camera on the Isaac Newton Telescope. The reported inclination of $31^{\circ}$ was calculated using equation \ref{eq:inclination} with assumed $Q_{\rm disk} = 0.2$.

Disk ellipticities were also measured by \citet{schmitt2000} in the same manner using $B$-band and $I$-band images from the Kitt Peak WIYN 0.9\,m telescope. The apparent axis ratios of $q_{{\rm disk,}\,B-\rm band} = 0.83$ and $q_{{\rm disk,}\,I-\rm band} = 0.844$ correspond to inclination angles 33--35$^\circ$ for the same assumed intrinsic thickness $Q_{\rm disk}=0.2$. Therefore, we conclude that the galaxy inclination is likely between $30^{\circ}$ and $40^{\circ}$.

\section{Dynamical Modeling}  \label{sec:dynamical_model}

We used the Schwarzschild orbit-superposition code FORSTAND \citep{vasiliev2020}, which is built on top of the AGAMA stellar-dynamical framework \citep{vasiliev2019}. The black hole mass is derived by generating self-consistent models that are constrained by the observed kinematics and photometry. While the NIFS spectroscopy characterizes the stellar motions near the central black hole, the SAURON spectrophotometric data help constrain the wider-field orbit structure and dark matter halo. Although FORSTAND is capable of building triaxial galaxy models that can simulate bars, the obvious axisymmetry of this low inclination lenticular galaxy necessitated the use of only axisymmetric models. 

The modelling procedure consists of several steps. For any choice of input parameters (disk thickness, SMBH mass, etc., except $\Upsilon$, which is varied separately as described below), the code first determines the deprojected 3D density profile and the corresponding gravitational potential, using the \texttt{CylSpline} Poisson solver. Then a large number of orbital initial conditions ($\sim 20\,000$) are randomly drawn from the stellar density profile, with their velocities assigned from an auxiliary Jeans model. With different random seeds, we may create several realizations of the orbit library to assess the impact of discreteness noise (finite number of orbits) on the $\chi^2$ values. The orbits are integrated in the given potential for 100 dynamical times, and the spatial density of each orbit is recorded on a grid in $R,z$. Its kinematic footprint is recorded on the intermediate 3D datacube, which is then convolved with the instrumental PSF and rebinned onto the Voronoi bins. The code then determines the orbit weights that minimize the deviation between the model and the observed kinematic datacubes, while satisfying the spatial density constraints exactly (in a discretized form).

Changing the overall mass normalization of the entire model (i.e., adjusting the stellar $\Upsilon$ in lockstep with the SMBH mass and the dark halo mass) is equivalent to rescaling the velocity axis of the model LOSVD by $\sqrt{\Upsilon}$, thus can be performed without reintegrating the orbits. Thus each orbit library is reused multiple times, scanning the range of $\Upsilon$ values in multiplicative steps of $2^{0.05}$, until the minimum of $\chi^2$ is found and bracketed from both ends. The construction of the orbit library takes $\sim 1$ minute, and each solution for the orbit weights requires $\sim 10$ seconds on a 32-core workstation. In total, we ran over a thousand realizations of orbit libraries, each one typically reused for $\sim 10$ values of $\Upsilon$.

We began our analysis by summarizing the free parameters of the model and their impact. Due to their considerable number, we did not explore the entire parameter space uniformly, but examined parameter combinations that were guided by our current understanding of the galaxy structure. 

\textbf{Surface Brightness Profile:} For the majority of our models, we relied on the $V$-band GALFIT surface brightness parametrization described in Section~\ref{sec:photometric_model} rather than the MGE method as described by \citet{cappellari2002}.  We also explored GALFIT density models from the $H$-band image, although it has a lower spatial resolution and is generally less reliable in separating the AGN from the galaxy nucleus; the constraints on $M_\bullet$ were similar in this case.

\textbf{Vertical Density Profile:} The vertical density profile of the disk can be represented by an exponential or $\rm sech^{2}$ function. In both cases, the total density profile is separable in radial and vertical directions, and the radial profile is exponential; the projected density contours have approximately constant ellipticity. Alternatively, the disk component may be represented by an oblate $n=1$ S{\'e}rsic profile, which is perfectly ellipsoidal with constant axis ratios in projection. Based on comparisons of the models using all three shapes, we chose the exponential profile for our main suite of models, but the results were similar for the other choices. The bulge is always represented by an ellipsoidal S{\'e}rsic profile.

\textbf{Stellar Mass-to-Light ratio ($\Upsilon$):} For simplicity, many stellar dynamical modeling studies employ a constant $\Upsilon$, although a few have investigated the effects of including a spatially-varying $\Upsilon$ \citep[e.g.,][]{mcconnell2013,nguyen2017}. Thus we began our exploration with models that assumed a single $\Upsilon$ across the galaxy.

As described above, the code automatically explores a range of $\Upsilon$ values and determines the best-fitting one, which usually tended to be on the lower end of the expected range of $2 - 4\,M_{\odot}/L_{\odot}$.

However, as explained in Section~\ref{sec:photometric_ML}, the galaxy color, and therefore $M/L$ of the stars, varies significantly over the footprint of the kinematic dataset. We therefore considered a series of models in which the bulge and disk $\Upsilon$ were different from each other, fixing the ratio $\Upsilon_{\rm disk}/\Upsilon_{\rm bulge}$ to a constant value $\mathcal R$ and allowing the code to vary $\Upsilon_{\rm bulge}$ to better match the kinematics. From the color gradient we expect $\mathcal R \sim 0.4$--$0.5$, but we varied $\mathcal R$ between 0.4 and 1.  

\textbf{SMBH mass:} This is our main quantity of interest, and it scales with $\Upsilon$ in each series of models sharing the same orbit library. We varied the ``baseline'' value (corresponding to $\Upsilon=1$) between zero and $2\times10^7\,M_\odot$ in steps of (1--2)$\times10^6\,M_\odot$. To avoid confusion, we always report the actual physical value (multiplied by $\Upsilon$) in the text and plots.

\textbf{Disk axis ratio ($Q_{\rm disk}$):} \citet{sandage1970} found that S0 galaxies likely exhibit average intrinsic disk flattening values of $0.25 \pm 0.06$, but the distribution includes values between 0.1--0.4. In some ways, NGC\,5273 is observationally more similar to a spiral galaxy than a lenticular galaxy as it has a more compact bulge than most S0 galaxies \citep[e.g.,][]{mendezabreu2008}. The generally adopted disk flattening value for standard spiral galaxies is 0.2 \citep[e.g.,][]{hubble1926,holmberg1958,tully2000}. Given the observed range of disk flattening in galaxies, we tested values between 0.1 and 0.4 in steps of 0.1. The inclination $i$ is computed from the apparent disk axis ratio $q_{\rm disk}$ and the assumed intrinsic axis ratio $Q_{\rm disk}$ (Equation~\ref{eq:inclination}).

\textbf{Bulge axis ratio ($Q_{\rm bulge}$):} From the GALFIT parameterization of the surface brightness profile, it is evident that the bulge is moderately oblate. By default, the bulge flattening is set by the observed flattening $q_{\rm bulge}$ and calculated inclination angle $i$ of the disk, but can be modified manually if desired, as described in Section~\ref{sec:inclination}. To investigate how this constraint affects the kinematic fits, we examined models with intrinsic bulge flattening values between $Q_{\rm disk}$ and 0.7 in steps of 0.1.

\textbf{Dark matter halo circular velocity($v_{\rm halo}$):} We assume a cored logarithmic potential of the dark halo with a core radius $r_{\rm halo}$ and asymptotic circular velocity $v_{\rm halo}$, which provides a circular-velocity profile $v_{\circ}(r) \equiv \sqrt{r\,\mathrm{d}\Phi/\mathrm{d}r} = v_{\rm halo} / \sqrt{1 + (r_{\rm halo}/r)^2}$. Given that the footprint of the SAURON kinematic dataset extends only to $\sim 1.4$~kpc, we would not be able to constrain both these parameters, so we fix $r_{\rm halo}$ to $20\arcsec$ (1.6~kpc) and explore different values of $v_{\rm halo}$. Note that the actual asymptotic circular velocity scales as $\sqrt{\Upsilon}$ in each series of models sharing the same orbit library. In what follows, we either quote the actual physical value of $v_{\rm halo}$ after rescaling, or write it as $\sqrt{\Upsilon}\,v_{\rm halo,1}$.

\textbf{Number of orbits:} We use 20\,000 orbits for most of our runs -- a sufficient number for the $\sim400$ density and $\sim1100$ kinematic constraints \citep{valluri2004}, but repeated a few runs increasing the orbit number to 100\,000, which reduces the best-fit $\chi^2$ by $\sim10-20$, but does not change the relative ranking of models with different parameters.

\textbf{Random seed:} Because the orbit library is constructed using randomly generated initial conditions, we ran several realizations of each model with different random seeds. 
Although the overall shape of the $\chi^2$ contours is similar, the precise location of the minimum and the associated value $\chi^2_{\rm min}$ vary between different realizations, with a typical scatter in $\chi^2$ at the level of 3--6 (i.e., considerably larger than the formal statistical significance level).

\section{Results}  \label{sec:results}

\begin{figure*}
\centering
\includegraphics{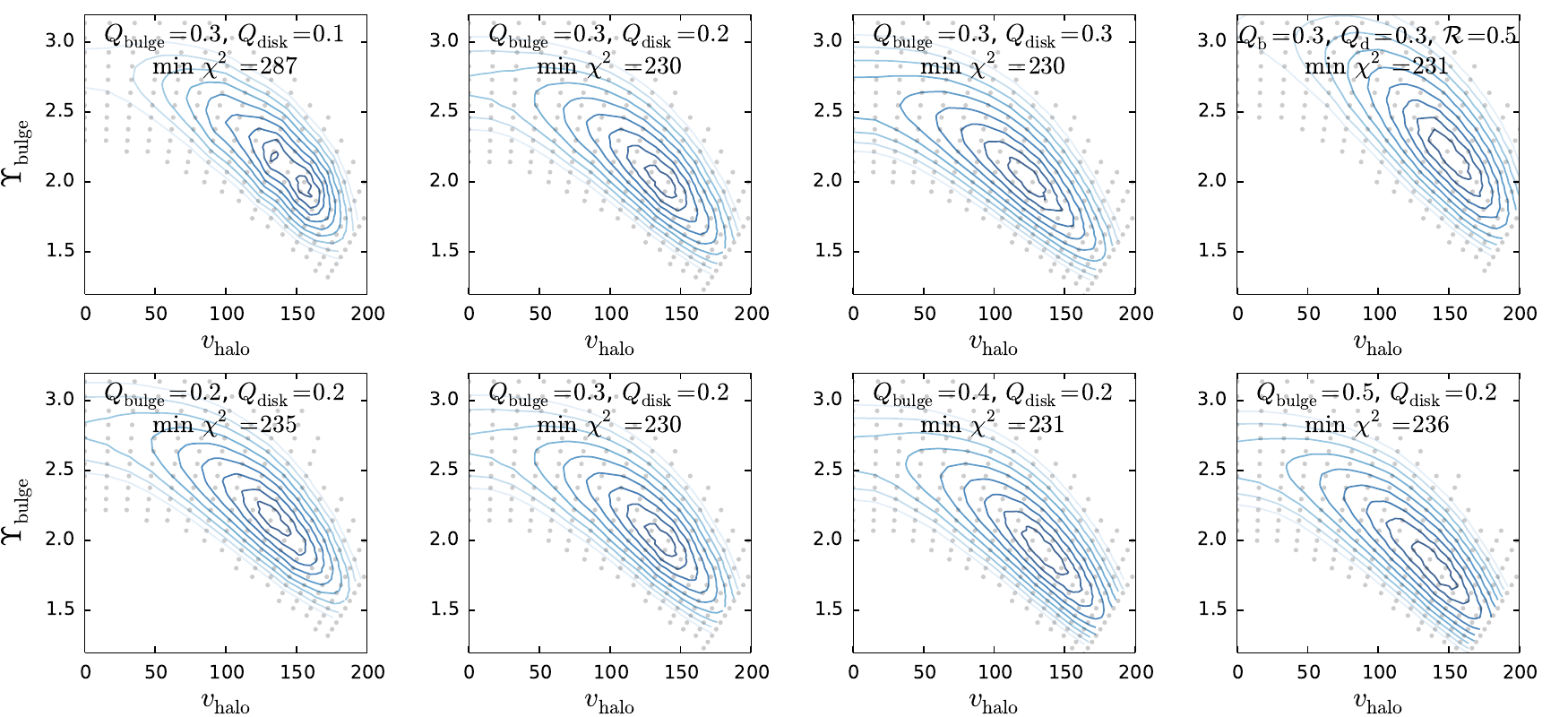}
\caption{Contours of $\Delta\chi^{2}\equiv\chi^{2}-\chi^{2}_{\rm min}$ as a function of $v_{\rm halo}$ ($x$-axis) and $\Upsilon_{\rm bulge}$ ($y$-axis) for several series of models constrained only by SAURON kinematics. Contours are placed at $\Delta\chi^2=2.3, 6.2, 11.8, \dots$, equivalent to $1\sigma$, $2\sigma$, $3\sigma$, \dots confidence intervals for two degrees of freedom. \protect\\
In the first row, $Q_{\rm bulge}$ is fixed to 0.3.  In the first three panels, $Q_{\rm disk}$ increases from 0.1 to 0.3 and $\mathcal R \equiv \Upsilon_{\rm disk}/\Upsilon_{\rm bulge} = 1.0$.  The last panel shows models with the same geometry as in the second panel, but a lower $\mathcal R = 0.5$, which naturally has a higher best-fit $\Upsilon_{\rm bulge}$.
In the second row, $Q_{\rm disk}$ is fixed to 0.2, $\mathcal R = 1.0$, and $Q_{\rm bulge}$ varies from 0.2 to 0.5 across columns (thus both panels in the second column are identical).
}
\label{fig:chi2_sauron}
\end{figure*}

Given the number of adjustable parameters in the models ($M_\bullet$, $\Upsilon_{\rm bulge}$, $\Upsilon_{\rm disk}$, $Q_{\rm bulge}$, $Q_{\rm disk}$, $v_{\rm halo}$, and the choice of vertical disk density profile), we did not attempt to explore all possible parameter combinations. Instead, we performed an initial coarse scan of the parameter space and identified the region of acceptable models, then explored the behaviour of models when varying one of the parameters while keeping the other fixed. A single series of models is defined by fixing $Q_{\rm bulge}$, $Q_{\rm disk}$, $v_{\rm halo}$, $\mathcal R\equiv \Upsilon_{\rm disk}/\Upsilon_{\rm bulge}$ and the disk vertical profile, and exploring a range of $M_\bullet$ values, with each orbit library used to generate a series of model fits with different $\Upsilon_{\rm bulge}$. 

\subsection{SAURON-only models}  \label{sec:results_sauron}

We begin by considering models constrained by SAURON data alone (only $v_{\rm LOS}$ and $\sigma_{\rm LOS}$). Naturally, given the low spatial resolution, these are insensitive to the SMBH mass, but can provide insights about the large-scale structure of the galaxy, namely the disk thickness and the contribution of the dark halo to the total potential. As mentioned above, we fix the halo core radius to $20\arcsec$, which makes its circular-velocity curve follow nearly the same radial profile as that of the disk. Therefore, we expect a large degree of degeneracy between $\Upsilon$ and $v_{\rm halo}$ when fitting the rotational velocity profile. However, the two parameters are not entirely degenerate, because making the disk more massive at the expense of the halo, while keeping the circular velocity unchanged, makes the total potential more strongly flattened. When the mass is more strongly confined to the equatorial plane, the vertical velocity dispersion $\sigma_z$ is also lower: in an isolated thin isothermal disk, $\sigma_z\propto \sqrt{h_{\rm disk}}$. Given the inclination of $\sim 35^\circ$, about 80\% of the vertical velocity contributes to the line-of-sight velocity. Therefore, to match the observed $\sigma_{\rm LOS}$ field, the models with a given $Q_{\rm disk}$ have to stay in a particular range of disk-to-halo mass ratios. Equivalently, if we shift the balance between disk and halo contributions to the circular-velocity curve towards larger $\Upsilon$ and lower $v_{\rm halo}$, these models have to become thinner (lower $Q_{\rm disk}$) to compensate for a stronger vertical force in the disk and keep the velocity dispersion in agreement with the observed value.

\begin{figure*}
\centering
\includegraphics{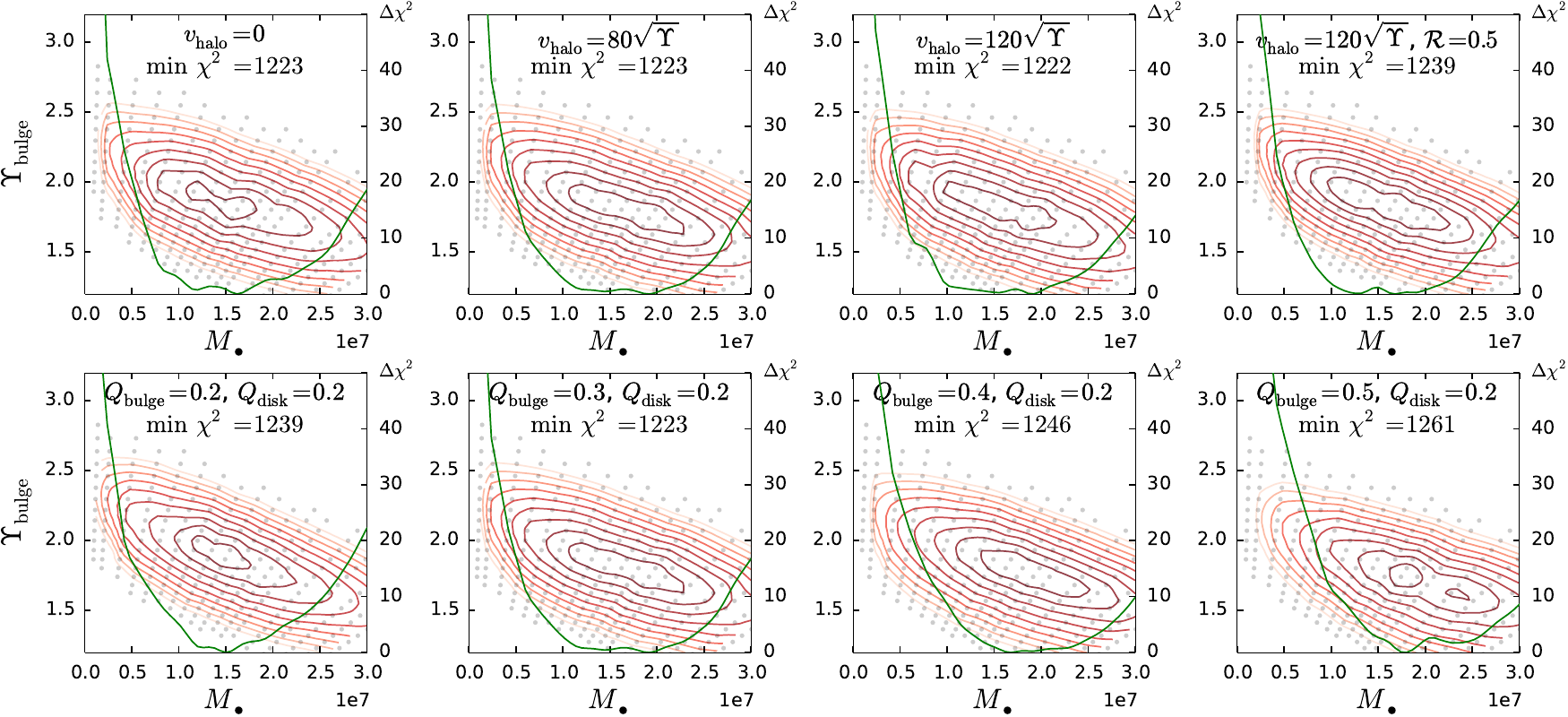}
\caption{Contours of $\Delta\chi^{2}\equiv\chi^{2}-\chi^{2}_{\rm min}$ as a function of $M_\bullet$ ($x$-axis) and $\Upsilon$ ($y$-axis) for several series of models constrained only by NIFS kinematics. 
The green lines are the marginalized $\Delta\chi^2$ values as a function of $M_\bullet$ alone, plotted in the range $\Delta\chi^2\le 50$ as shown in the secondary $y$-axis. \protect\\
In the first row, we fix $Q_{\rm disk}=0.2$ and $Q_{\rm bulge}=0.3$. We set $\Upsilon_{\rm disk}=\Upsilon_{\rm bulge}$ and vary $v_{\rm halo}$ from 0 to 120$\sqrt{\Upsilon}\,\mathrm{km\:s}^{-1}$ in the first three panels.  In the last panel, we set $\Upsilon_{\rm disk}=0.5\Upsilon_{\rm bulge}$ (in other panels both components have the same $\Upsilon$).
In the second row, we fix $v_{\rm halo}=80\sqrt{\Upsilon}\,\mathrm{km\:s}^{-1}$ and $Q_{\rm disk}=0.2$ and vary $Q_{\rm bulge}$ from 0.2 to 0.5 across columns. Both panels in the second column are identical.
}  \label{fig:chi2_nifs}
\end{figure*}

Figure~\ref{fig:chi2_sauron} confirms these expectations. In a series of models with different disk thickness (top row), the best-fit $\Upsilon$ lies in the range 2.5--3.0 in absense of dark halo, but decreases to $2.0$--$2.5$ for models with the optimal choice of $v_{\rm halo} \simeq (80$--$100)\sqrt{\Upsilon}~\mathrm{km\:s}^{-1}$. Very thin disks are discouraged by the data, being unable to produce high enough velocity dispersion without exceeding the observed rotational velocity, but values of $Q_{\rm disk}=0.2$--0.4 seem reasonable. The bulge axis ratio has a secondary effect on the fit quality (bottom row), but values of $Q_{\rm bulge}\gtrsim 0.5$ produce worse fits. On the other hand, we may reasonably expect that the bulge would not be thinner than the disk (both from theoretical arguments and based on the rounder shape of isophotes in the inner part), thus impose a lower limit $Q_{\rm bulge}\ge Q_{\rm disk}$. There was little quantitative difference between models with different choices of vertical disk density profile, with the exponential disk being slightly better overall. Finally, models with $\Upsilon_{\rm disk} < \Upsilon_{\rm bulge}$ have necessarily higher best-fit values of $\Upsilon_{\rm bulge}$ and a higher contribution of dark halo (top right panel).

\subsection{NIFS-only models}  \label{sec:results_nifs}

We then turn our attention to the inner galaxy region, constructing models constrained by NIFS kinematics alone. Since the bulge density exceeds the disk density by at least an order of magnitude within the NIFS footprint (Figure~\ref{fig:density_profile}), these models are expected to be insensitive to the disk axis ratio $Q_{\rm disk}$ or the dark halo velocity $v_{\rm halo}$, so we usually fix these to the best-fit values determined in the previous section, namely $Q_{\rm disk}=0.2$ and $v_{\rm halo}=80\sqrt{\Upsilon}~\mathrm{km\:s}^{-1}$, but explore the effect of variation of these parameters in one series of models. The remaining free parameters are $M_\bullet$, $Q_{\rm bulge}$, and of course $\Upsilon_{\rm bulge}$. 

Figure~\ref{fig:chi2_nifs} illustrates models with different dark halo velocity (first row) or different bulge flattening (second row).
Similarly to the case of SAURON data, there exists a certain degeneracy between parameters controlling the two contributions to the total potential, namely the SMBH mass and stellar $M/L$. Increasing $M_\bullet$ and simultaneously lowering $\Upsilon$ keeps the overall gravitational field roughly constant, but of course, its spatial variation is different between the two components, and therefore the range of acceptable values of $M_\bullet$ is also limited (primarily by the central velocity dispersion peak). The profiles of $\chi^2$ as a function of $M_\bullet$ generally have a broad minimum in the range (0.5--2.5)${}\times 10^7\,M_\odot$ and rise steeply towards smaller $M_\bullet$ and less steeply towards larger values. The corresponding range of $\Upsilon$ values is 1.6--2.0, somewhat lower than in the SAURON-only models. As expected, dark halo has little effect on the NIFS kinematics (first row), while making $\Upsilon_{\rm disk}<\Upsilon_{\rm bulge}$ (last panel) marginally deteriorates the fit. However, in order for $\Upsilon_{\rm disk}$ to be compatible with the SAURON-only models (Figure~\ref{fig:chi2_sauron}), $v_{\rm halo}$ must be $\gtrsim 80\sqrt{\Upsilon}~\mathrm{km\:s}^{-1}$.

The second row of Figure~\ref{fig:chi2_nifs} shows that more strongly flattened models (lower $Q_{\rm bulge}$) need to have a higher $\Upsilon$ to reproduce the observed velocity dispersion outside the SMBH radius of influence, although the effect is much smaller than in the case of large-scale disk. We again find that the preferred range of bulge axis ratio is on the lower side ($Q_{\rm bulge} \simeq 0.3$) compared to the expected values from the GALFIT photometric model ($Q_{\rm bulge} \simeq 0.6$--0.7). However, the difference in the projected axis ratio $q_{\rm bulge}$ is only a few per cent, and the low intrinsic axis ratio is not implausible a priori.

\subsection{Models fitted to both kinematic datasets}  \label{sec:results_both}

\begin{figure*}
\centering
\includegraphics{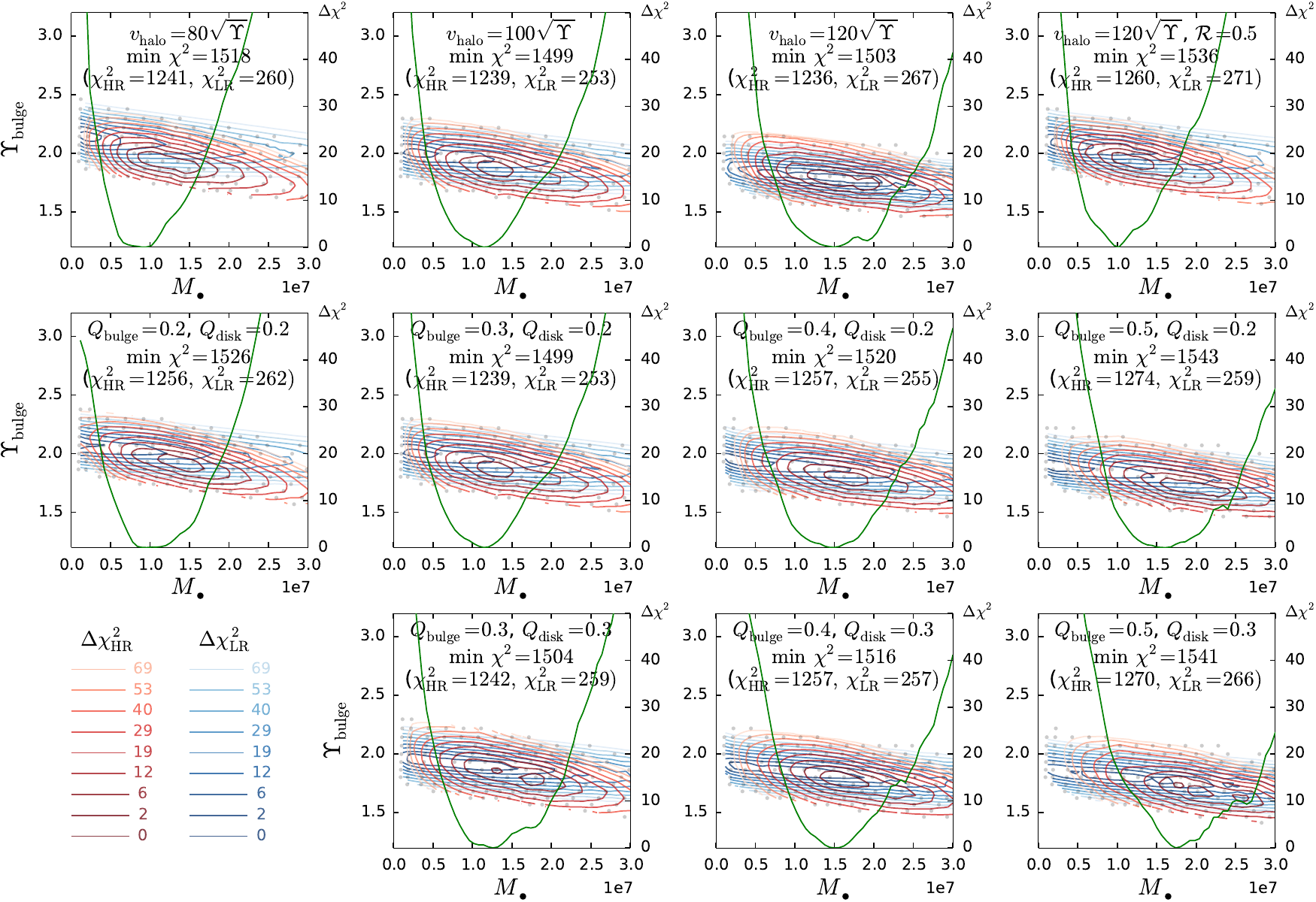}
\caption{Contours of $\Delta\chi^{2}\equiv\chi^{2}-\chi^{2}_{\rm min}$ as a function of $M_\bullet$ ($x$-axis) and $\Upsilon_{\rm bulge}$ ($y$-axis) for various choices of other parameters.
Values of $\Delta\chi^2$ are plotted separately for NIFS (red) and SAURON (blue) datasets, but the models are fitted to both datasets simultaneously. The green lines show the total marginalized $\Delta\chi^{2}$ contours as a function of black hole mass, and the lowest value for each panel is shown in the caption.
The first two rows have $Q_{\rm disk}=0.2$ and the last row -- $Q_{\rm disk}=0.3$. In the first row, we fix $Q_{\rm bulge}=0.3$ and vary $v_{\rm halo}$ from 80$\sqrt{\Upsilon}$ to 120$\sqrt{\Upsilon}$ $\mathrm{km\:s}^{-1}$ in the first three columns; the last column additionally has a lower $\Upsilon_{\rm disk} = 0.5\,\Upsilon_{\rm bulge}$ (in all other panels these two parameters are equal). In the second and third row, we fix $v_{\rm halo}=100\sqrt{\Upsilon}\,\mathrm{km\:s}^{-1}$ and $Q_{\rm disk}$ to 0.2 or 0.3 respectively, and increase $Q_{\rm bulge}$ from $Q_{\rm disk}$ to 0.5 across columns. The overall lowest $\chi^2$ is attained in the panels in the second column, first or second row (which are identical), but the $\chi^2$ profiles as a function of $M_\bullet$ are qualitatively similar across all panels and have minima in the range $M_\bullet=(0.5$--$2.0)\times 10^7\,M_\odot$.}
\label{fig:chi2_both}
\end{figure*}

Finally, we are in the position to explore the models fitted to both datasets simultaneously. Guided by the earlier analysis, we limit the range of $v_{\rm halo}$ to (80--120)$\sqrt{\Upsilon}~\mathrm{km\:s}^{-1}$, $Q_{\rm disk}$ to 0.2--0.4, let $Q_{\rm bulge}$ vary between $Q_{\rm disk}$ and 0.5, and run a few model series with $\Upsilon_{\rm disk}/\Upsilon_{\rm bulge}<1$. 

Figure~\ref{fig:chi2_both} shows a collection of models with different choices of parameters.
The contours of $\Delta\chi^{2}\equiv\chi^{2}-\chi^{2}_{\rm min}$ are displayed in each panel as a function of $M/L$ ratio $\Upsilon_{\rm bulge}$ on the y-axis and black hole mass $M_\bullet$ on the x-axis. 
The models are optimized for the total $\chi^2$ score (the sum of contributions from NIFS and SAURON), but we opted to display the $\chi^2$ values from both datasets separately (NIFS $\chi^2_{\rm HR}$ in red and SAURON $\chi^2_{\rm LR}$ in blue), while the sum is not shown to avoid clutter. Instead, we plot the 1d $\Delta\chi^2$ profiles as a function of $M_\bullet$ alone (green lines), which are obtained by marginalizing the likelihoods over the $\Upsilon$ axis.

Given that the $\Upsilon_{\rm bulge}$ preferred by the NIFS-only models is somewhat lower than for the SAURON-only models, the combined models attain lowest $\chi^2$ values in an intermediate range of $\Upsilon$, and the overall $\chi^2$ score is higher if there is a larger difference in the $\Upsilon$ values preferred by each dataset.
Since the SAURON kinematic fits are insensitive to $M_\bullet$, the contours of $\chi^2_{\rm LR}$ are nearly horizontal in the $M_\bullet$--$\Upsilon$ plane, while those of $\chi^2_{\rm HR}$ are tilted diagonally. The best overall fit is achieved where the two ``degeneracy valleys'' intersect each other, but because the NIFS contours are tilted only slightly, the marginalized $\chi^2$ profiles as a function of $M_\bullet$ still have rather broad minima in the range $M_\bullet=(0.5$--$2.0)\times10^7\,M_\odot$, and are qualitatively similar among all series of models, with the lowest overall $\chi^2$ achieved for $Q_{\rm bulge}=0.3$.

The best-fit values of $\Upsilon_{\rm bulge}$ are slightly below 2.0, and models with $\Upsilon_{\rm disk}=0.5\Upsilon_{\rm bulge}$ are again disfavoured by the data, contrary to expectations from the color gradient. The $M/L$ from dynamical models is lower than the photometric estimate (2.0--4.0, Section~\ref{sec:photometric_ML}), but this is achieved by adding a rather massive dark halo, which dominates the circular-velocity curve at large distances (at the edge of the SAURON footprint). As discussed in Section~\ref{sec:results_sauron}, one may adjust the balance between stars and dark halo to some extent, while keeping the circular-velocity curve (and hence the mean rotation velocity) unchanged. However, models with higher $\Upsilon_{\rm disk}$ and lower $v_{\rm halo}$ are incompatible with NIFS data, which prefer $\Upsilon_{\rm bulge}\lesssim 2$, and since we do not expect $\Upsilon_{\rm disk}$ to be higher than $\Upsilon_{\rm bulge}$, the optimal choice is to make them equal. 

\begin{figure*}
\centering
\includegraphics{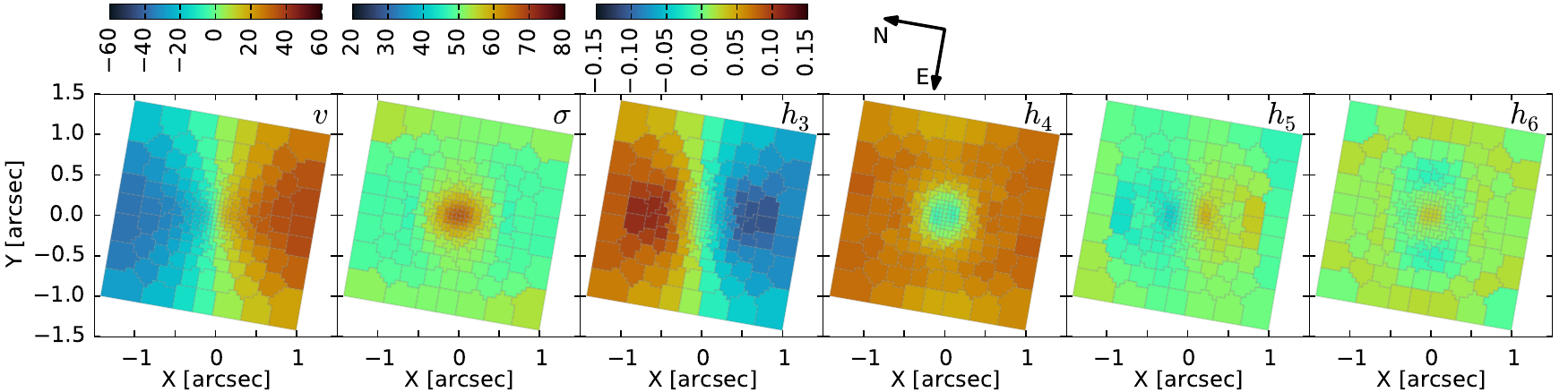}
\caption{Stellar kinematic maps of the fiducial model for the NIFS dataset, which can be compared to the observational maps (Figure~\ref{fig:nifs_maps}). This model has an exponential vertical disk profile (Equation~\ref{eq:disk_vertical_profile}) with $h_{\rm disk}/R_{\rm disk}=0.2$, a flattened bulge with $Q_{\rm bulge}=0.3$, stellar $M/L$ $\Upsilon=1.9$, a dark halo with $v_{\rm halo}=100\sqrt{\Upsilon}~\mathrm{km\:s}^{-1}$, and a black hole $M_\bullet=10^7\,M_\odot$.
}
\label{fig:nifs_maps_model}
\end{figure*}

\begin{figure}
\centering
\includegraphics{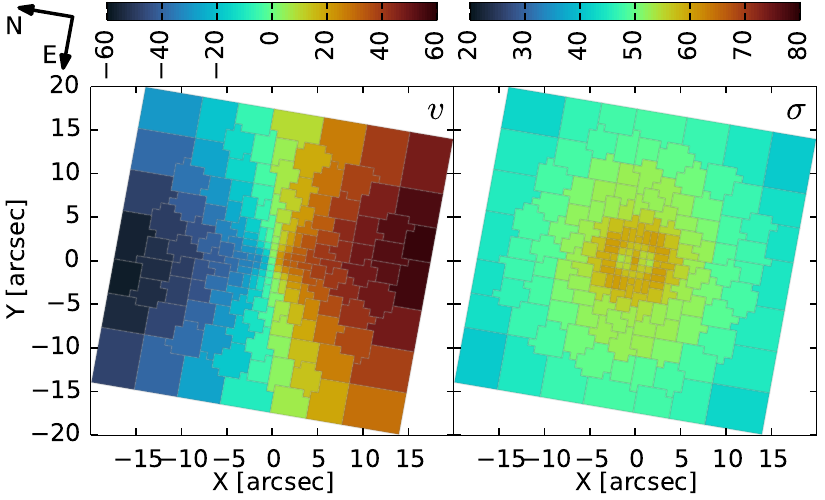}
\caption{Stellar kinematic maps of the fiducial model (same as in Figure~\ref{fig:nifs_maps_model}) for the SAURON dataset, which can be compared to the observational maps (Figure~\ref{fig:sauron_maps}).
}
\label{fig:sauron_maps_model}
\end{figure}
 
\begin{figure}
\centering
\includegraphics{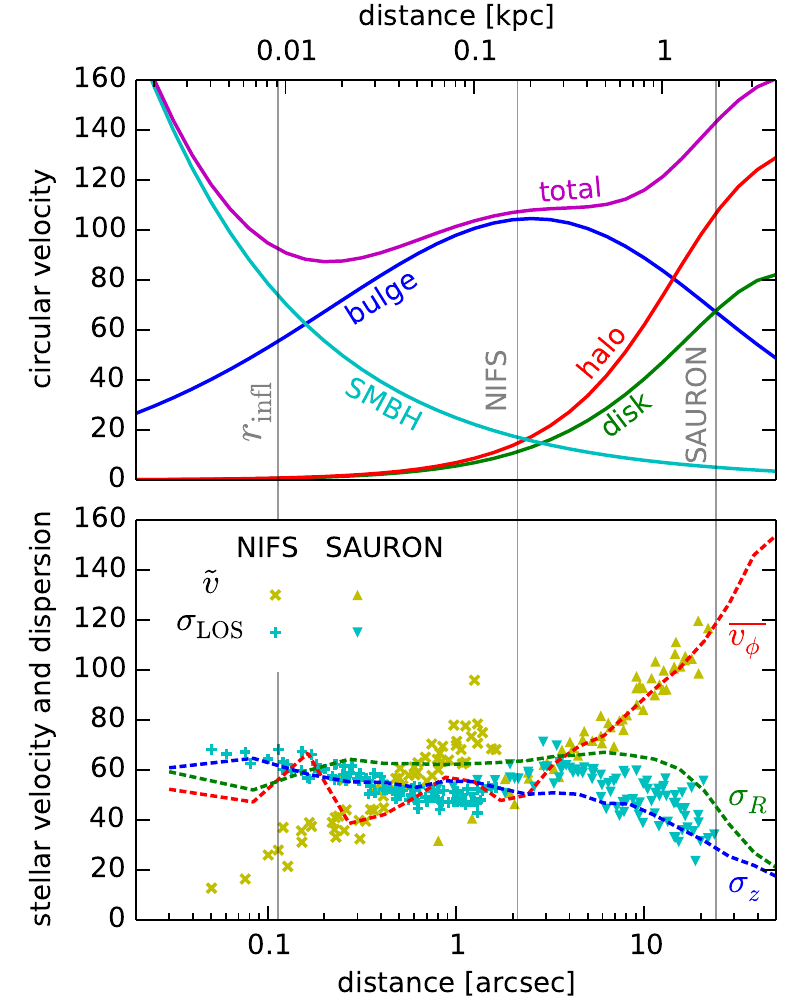}
\caption{Properties of the fiducial model (same as in Figure~\ref{fig:nifs_maps_model}).
Top panel shows the circular velocity $v_{\rm circ}(R)\equiv \sqrt{R\,\mathrm{d}\Phi/\mathrm{d}R}$ for each potential component (SMBH, bulge, disk, halo) and the total one. The bottom panel shows the the mean rotational velocity (red dashed curve), radial (green dashed curve) and vertical (blue dashed curve) velocity dispersions of the orbit-superposition model as functions of galactocentric radius, averaged over the vertical direction. For comparison, yellow and cyan symbols plot the observed velocity and dispersion from NIFS (crosses and pluses) and SAURON (upward and downward triangles). To make the comparison more meaningful, the position of each Voronoi bin in the sky plane ($x, y$, where $x$ is the coordinate along the major axis) is converted to the galactocentric distance as $R = \sqrt{x^2 + (y/\cos i)^2}$, and the line-of-sight velocity is translated into rotational velocity as $\tilde v = v_{\rm LOS}\,R / (x\, \sin i)$; these expressions would produce an exact deprojection in case of a thin disk observed at an inclination $i$.
}
\label{fig:rotcurve}
\end{figure}

Due to the degeneracy between $\Upsilon_{\rm bulge}$ and $M_\bullet$, the location of the minimum in $\chi^2$ as a function of $M_\bullet$ depends on parameters that affect $\Upsilon$, namely $v_{\rm halo}$ and $Q_{\rm disk/bulge}$. The impact of dark halo on the inferred $M_\bullet$ has been discussed by, for example, \citet{gebhardt2009,erwin2018}. Nevertheless, this degeneracy valley has a rather sharp edge at the lower end ($\chi^2$ rapidly rises for $M_\bullet \le 5\times10^6\,M_\odot$ in all model series), and while it only gradually disappears beyond $2\times 10^7\,M_\odot$, these higher values of $M_\bullet$ correspond to increasingly lower $\Upsilon$ that can only be sustained by an unrealistically high $v_{\rm halo}$ and are incompatible with photometric estimates of $M/L$. We thus conclude that the SMBH mass likely lies in the range (0.5--$2.0)\times10^7\,M_\odot$ for all plausible combinations of other parameters, but are unable to place stronger constraints on it.

The values of $\chi^2$ in the best-fit models are higher than could be expected from the random noise in the observations: namely, $\chi^2_{LR}\simeq 250$ for $2\times 91$ kinematic constraints in the SAURON dataset ($v$ and $\sigma$ in 91 Voronoi bins), and $\chi^2_{HR}\simeq 1240$ for $6\times 97$ kinematic constraints in the NIFS dataset. Some of the discrepancy could be attributed to the underestimation of observational uncertainties, but there are also systematic deviations across the NIFS field, most likely associated with a slight misalignment between photometric and kinematic major axes and a somewhat elongated shape of the central velocity dispersion peak in the data, as opposed to a nearly-circular peak in the model. Both features could be indicative of a more complicated kinematics of the central region than assumed in our axisymmetric models, possibly a signature of a nuclear bar or a misaligned nuclear disk. Figures~\ref{fig:nifs_maps_model} and \ref{fig:sauron_maps_model} show the NIFS and SAURON kinematic maps of the fiducial model, which do reproduce most of the important features in the observations, including the central velocity peak, anticorrelation between $v$ and $h_3$, and a positive $h_4$ outside the central peak. 

Figure~\ref{fig:rotcurve} shows the gravitational field of the fiducial model represented by the circular-velocity curves of several components (bulge, disk, halo and the SMBH). As discussed above, if the stellar $M/L$ is constant across the galaxy, it ends up being rather low ($\Upsilon \lesssim 2$) and one needs a massive dark halo to sustain the rotational velocity in the disk. Making $\Upsilon_{\rm disk} < \Upsilon_{\rm bulge}$ as suggested by the observed color gradient would only exacerbate the problem: since $\Upsilon_{\rm bulge}$ is largely fixed by the NIFS kinematics, $\Upsilon_{\rm disk}$ would be even lower in the outer region, and must be compensated by an even more massive halo. There is also a noticeable mismatch between the amplitudes of the deprojected rotational velocity at the outer edge of the NIFS dataset (yellow crosses in the bottom panel) and the SAURON data in the same spatial region, but it is entirely explained by the larger PSF of the SAURON instrument, which smoothes out the velocity gradient. The rotational velocity in the model remains quite high down to the very center ($\overline{v_\phi}/\sigma \simeq 1$), confirming the designation of this galaxy as a fast rotator \citep{krajnovic2013}.\\

\section{Discussion}

\subsection{Mass-to-Light Ratio}

Our $V$-band $M/L$ values estimated from the FORSTAND $\chi^{2}$ contour plots ($\sim 1.8 - 2.0\,M_{\odot}/L_{\odot}$) are on the low side of the expectations from the galaxy colors ($\sim 2 - 4\,M_{\odot}/L_{\odot}$). We also investigated models using a lower-resolution ground-based $H$-band image of the galaxy with a wider FOV, but again found that the best-fit $M/L$ values were lower than expected from the photometry, even in the $H$-band.

Since the NIFS FOV mostly encompasses the bulge and the SAURON FOV mainly includes the light from the disk, each is likely to be dominated by a different stellar population. By forcing the two data sets to match, we may be striving for lower $\chi^{2}$ minima at the expense of results that reflect physical properties of the galaxy. The stellar distribution and structure of spiral galaxies changes gradually with radius, but are especially different between the bulge and disk. However, we did not find a significantly different value of $M/L$ when considering the NIFS-only models, which should be dominated by the bulge and a more consistent stellar population. In future studies, it may be beneficial to explore the results obtained with a $M/L$ gradient. Because we adopted a single stellar template when fitting the kinematics for each dataset, the kinematics would need to be re-derived as well.

\subsection{Black Hole Mass}

In response to the low $M/L$ values, the stellar dynamical models find black hole masses between $M_{\bullet} = [0.5 - 2.0] \times 10^{7}$\,$M_{\odot}$, which just agree on the low end with the reverberation mass of $M_{\bullet} = [4.7 \pm 1.6] \times 10^{6}$\,$M_{\odot}$ \citep{bentz2014}. It is possible that the population-average $\langle f \rangle$ factor (used to scale reverberation masses so they are in general agreement with the $M_{\bullet}-\sigma$ relationship) applied to the reverberation mass is not the right value for this particular galaxy. Though a common approach in reverberation mapping is to use a mean $\langle f \rangle$ factor based on measurements from many galaxies, each AGN has a unique value that is influenced by the inclination of the accretion disk to our line of sight, which is not necessarily correlated with the inclination of the galaxy \citep[e.g.,][]{gallimore2006}. A more face-on orientation would require a larger scale factor because a smaller component of the broad-line region velocity would be visible along our line of sight. A larger $f$ factor, in turn, produces a larger reverberation mass, and could bring the reverberation mass into even better agreement with the stellar dynamical mass. 

However, studies of NGC\,5273 have shown that the orientation of the AGN is probably viewed at an intermediate angle and is not extremely face-on. A time-lag analysis of X-ray observations \citep{vincentelli2020} indicated that the accretion disk is viewed at an inclination $\geq 45^{\circ}$. \citet{ulvestad1984} found partially extended radio emission, also suggesting that the AGN is not face-on. An X-ray spectral analysis of NGC\,5273 found that reflection models preferred ``low" inclinations between $28^{\circ} - 50^{\circ}$ \citep{pahari2017}. All of these estimates of the inclination of the AGN in NGC\,5273 suggest that the adopted $\langle f \rangle = 4.3$ \citep{grier2013}, which suggests a population average inclination of $\sim 30^{\circ}$, is not egregiously erroneous.

The best-fit black hole masses from our stellar dynamical models are also somewhat larger than expected based on the $M_{\bullet}-\sigma$ relationship. Using the formulation of \citet{vandenbosch2016} with a bulge stellar velocity dispersion of $\sigma_{R_{e}/8} = 74.1 \pm 3.7\,\mathrm{km\:s^{-1}}$ \citep{cappellari2013b} predicts $M_{\bullet} = [1.03 \pm 0.37] \times 10^{6}$\,$M_{\odot}$. While on the high side, our best-fit range of black hole mass does fall within the 0.49\,dex scatter of the $M_{\bullet}-\sigma$ relation shown in Figure 1 of \citet{vandenbosch2016}.

\subsection{Black Hole Mass Comparisons}

Comparisons of black hole masses derived from multiple independent techniques are vital to identifying biases in the methods given their inherently different assumptions.  NGC\,5273 is only the third galaxy with a black hole mass measurement from reverberation mapping and from stellar dynamical modeling, and we find moderate agreement between the results of the two methods.  This is in agreement with the findings for NGC\,4151, but at odds with the results for NGC\,3227, as we summarize below.

The central black hole in NGC\,4151 has reverberation mapping and stellar dynamical modeling masses that agree quite well. The reverberation mass of $M_{\bullet} = [1.66^{+0.48}_{-0.34}] \times 10^{7}$\,$M_{\odot}$ \citep{bentz2022} was recently re-derived from spectroscopic data collected in 2005 \citep{bentz2006}. This new analysis did not rely on assuming a specific value of $\langle f \rangle$ because the velocity-resolved emission-line response was instead modeled to constrain the inclination, geometry, and kinematics of the broad line region gas directly. The stellar dynamical modeling mass of $M_{\bullet} = [0.25 - 3] \times 10^{7}$\,$M_{\odot}$ \citep{roberts2021} was re-measured from observations obtained by \citet{onken2014}, with several improvements to the image reduction process and utilizing the FORSTAND dynamical modeling algorithm. Furthermore, it relies on an accurate distance of $D = 15.8 \pm 0.4$\,Mpc from analysis of Cepheid variable stars \citep{yuan2020}. Finally, NGC\,4151 has a gas dynamical modeling mass \citep{hicks2008} that, when scaled to the Cepheid distance, is $M_{\bullet} = [3.6^{+0.9}_{-2.6}] \times 10^{7}$\,$M_{\odot}$ and agrees well with the reverberation mapping and stellar dynamical modeling masses.

On the other hand, there is less agreement between the masses measured for NGC\,3227. Reverberation masses (all scaled to assume $\langle f \rangle = 4.3$) of $M_{\bullet} = [3.6 \pm 0.4] \times 10^{6}$\,$M_{\odot}$ and $M_{\bullet} = [4.4 \pm 2.2] \times 10^{6}$\,$M_{\odot}$ based on monitoring campaigns in two separate years are reported by \citet{derosa2018}, and a mass of $M_{\bullet} = [6.7 \pm 1.4] \times 10^{6}$\,$M_{\odot}$ is reported by \citet{denney2010}. The most accurate distance to NGC\,3227 is $D = 23.7 \pm 2.6$\,Mpc and is derived from analysis of the surface brightness fluctuations of NGC\,3226 \citep{tonry2001}, with which NGC\,3227 is interacting. Scaling the stellar dynamical modeling mass \citep{davies2006} to this adopted distance gives $M_{\bullet} = [19 \pm 9] \times 10^{6}$\,$M_{\odot}$. A gas dynamical modeling mass has also been published for NGC\,3227 \citep{hicks2008}, and scaling that mass to the same distance gives $M_{\bullet} = [30^{+15}_{-6}] \times 10^{6}$\,$M_{\odot}$. Thus, while the dynamical masses agree, they are a factor of $4-5$ larger than the reverberation masses. Potentially exacerbating this disagreement, the dynamical modeling studies did not include dark matter, and thus may be underestimating the mass \citep[see for example][]{gebhardt2009}. A velocity-resolved analysis of the reverberation response in NGC\,3227 is currently underway and will avoid the introduction of an $\langle f \rangle$ factor in the mass (Robinson et al., in prep). This may help to explain some of the disagreement, although a reanalysis of the stellar dynamics in NGC\,3227 is also needed.

Additional tests of the agreement, or lack thereof, between black hole masses derived from stellar dynamical modeling and from reverberation mapping are needed to assess whether the black hole masses that are measured for local galaxies are all on the same mass scale.  Such tests are critical since results derived for local galaxies are often used to estimate black hole masses at cosmological distances (e.g., \citealt{kozlowski2017,wang2021}) and for studies of black hole and galaxy evolution (e.g., \citealt{heckman2014} and references therein). 

\section{Summary}

We have constrained the mass of the SMBH in NGC\,5273 using dynamical models of the stellar kinematics. We obtained AO-assisted observations of the nuclear stellar kinematics with Gemini NIFS, and reduced the observations with a new NIFS image reduction pipeline that improves upon the existing IRAF reduction scripts. To constrain the stellar kinematics on larger scales, we used observations from SAURON collected by the ATLAS$^{\textrm{3D}}$ collaboration. We adaptively binned the NIFS and SAURON data cubes and extracted the stellar kinematics by simultaneously fitting point-symmetric bins to determine the LOSVD. The surface brightness profile of the galaxy was decomposed into analytic profiles with Galfit, which were then deprojected to determine the 3D luminosity density. We used a new Schwarzschild orbit-superposition algorithm called FORSTAND to simulate the stellar orbits within the FOV and explore the effects of different assumptions for a variety of modeling parameters including stellar mass-to-light ratio, black hole mass, disk and bulge shape, and dark matter contribution. The range of acceptable models includes $M_{\bullet} = [0.5 - 2.0] \times 10^{7}$\,$M_{\odot}$ which agrees on the low end with the previously published reverberation mass. NGC\,5273 is only the third SMBH with mass constraints from both reverberation mapping and stellar dynamical modeling, and thus represents an important check on the accuracy of the most widely used black hole mass measurement techniques.  


KAM and MCB are supported by the NSF through grant AST-2009230 to Georgia State University. MV is supported by the NSF through grant AST-2009122. CAO was supported by the Australian Research Council (ARC) through Discovery Project DP190100252.

Based on observations obtained at the Gemini Observatory, a program of NSF's NOIRLab, which is managed by the Association of Universities for Research in Astronomy (AURA) under a cooperative agreement with the National Science Foundation on behalf of the Gemini Observatory partnership: the National Science Foundation (United States), National Research Council (Canada), Agencia Nacional de Investigaci\'{o}n y Desarrollo (Chile), Ministerio de Ciencia, Technolog\'{i}a e Innovaci\'{o}n (Argentia), Minist\'{e}rio da Ci\^{e}ncia, Tecnologia, Inova\c{c}\~{o}es e Comunica\c{c}\~{o}es (Brazil), and Korea Astronomy and Space Science Institute (Republic of Korea). This work was enabled by observations from the Gemini North telescope, located within the Maunakea Science Reserve and adjacent to the summit of Maunakea. We are grateful for the privilege of observing the Universe from a place that is unique in both its astronomical quality and its cultural significance. 


\facilities{Gemini (NIFS), WHT (SAURON), HST (WFPC2), APO (NIC-FPS)}

\software{pPXF (\citealt{cappellari2004,cappellari2017}), FORSTAND \citep{vasiliev2020},  NIFS reduction pipeline (https://github.com/remcovandenbosch/NIFS-pipeline), MGE \citep{cappellari2002}, GALFIT (\citealt{peng2002,peng2010})}

\bibliography{ngc5273sources}

\begin{thebibliography}{}
\expandafter\ifx\csname natexlab\endcsname\relax\def\natexlab#1{#1}\fi
\providecommand{\url}[1]{\href{#1}{#1}}
\providecommand{\dodoi}[1]{doi:~\href{http://doi.org/#1}{\nolinkurl{#1}}}
\providecommand{\doeprint}[1]{\href{http://ascl.net/#1}{\nolinkurl{http://ascl.net/#1}}}
\providecommand{\doarXiv}[1]{\href{https://arxiv.org/abs/#1}{\nolinkurl{https://arxiv.org/abs/#1}}}

\bibitem[{{Bacon} {et~al.}(2001){Bacon}, {Copin}, {Monnet}, {Miller},
  {Allington-Smith}, {Bureau}, {Carollo}, {Davies}, {Emsellem}, {Kuntschner},
  {Peletier}, {Verolme}, \& {de Zeeuw}}]{bacon2001}
{Bacon}, R., {Copin}, Y., {Monnet}, G., {et~al.} 2001, \mnras, 326, 23,
  \dodoi{10.1046/j.1365-8711.2001.04612.x}

\bibitem[{{Bambi}(2018)}]{bambi2018}
{Bambi}, C. 2018, Annalen der Physik, 530, 1700430,
  \dodoi{10.1002/andp.201700430}

\bibitem[{{Barth} {et~al.}(2016){Barth}, {Boizelle}, {Darling}, {Baker},
  {Buote}, {Ho}, \& {Walsh}}]{barth2016}
{Barth}, A.~J., {Boizelle}, B.~D., {Darling}, J., {et~al.} 2016, \apjl, 822,
  L28, \dodoi{10.3847/2041-8205/822/2/L28}

\bibitem[{{Barway} {et~al.}(2005){Barway}, {Mayya}, {Kembhavi}, \&
  {Pandey}}]{barway2005}
{Barway}, S., {Mayya}, Y.~D., {Kembhavi}, A.~K., \& {Pandey}, S.~K. 2005, \aj,
  129, 630, \dodoi{10.1086/426906}

\bibitem[{{Beifiori} {et~al.}(2011){Beifiori}, {Maraston}, {Thomas}, \&
  {Johansson}}]{beifiori2011}
{Beifiori}, A., {Maraston}, C., {Thomas}, D., \& {Johansson}, J. 2011, \aap,
  531, A109, \dodoi{10.1051/0004-6361/201016323}

\bibitem[{{Bell} \& {de Jong}(2001)}]{bell2001}
{Bell}, E.~F., \& {de Jong}, R.~S. 2001, \apj, 550, 212, \dodoi{10.1086/319728}

\bibitem[{{Bentz} {et~al.}(2022){Bentz}, {Williams}, \& {Treu}}]{bentz2022}
{Bentz}, M.~C., {Williams}, P.~R., \& {Treu}, T. 2022, \apj, 934, 168,
  \dodoi{10.3847/1538-4357/ac7c0a}

\bibitem[{{Bentz} {et~al.}(2006){Bentz}, {Denney}, {Cackett}, {Dietrich},
  {Fogel}, {Ghosh}, {Horne}, {Kuehn}, {Minezaki}, {Onken}, {Peterson}, {Pogge},
  {Pronik}, {Richstone}, {Sergeev}, {Vestergaard}, {Walker}, \&
  {Yoshii}}]{bentz2006}
{Bentz}, M.~C., {Denney}, K.~D., {Cackett}, E.~M., {et~al.} 2006, \apj, 651,
  775, \dodoi{10.1086/507417}

\bibitem[{{Bentz} {et~al.}(2014){Bentz}, {Horenstein}, {Bazhaw},
  {Manne-Nicholas}, {Ou-Yang}, {Anderson}, {Jones}, {Norris}, {Parks},
  {Saylor}, {Teems}, \& {Turner}}]{bentz2014}
{Bentz}, M.~C., {Horenstein}, D., {Bazhaw}, C., {et~al.} 2014, \apj, 796, 8,
  \dodoi{10.1088/0004-637X/796/1/8}

\bibitem[{{Blakeslee} {et~al.}(2001){Blakeslee}, {Lucey}, {Barris}, {Hudson},
  \& {Tonry}}]{blakeslee2001}
{Blakeslee}, J.~P., {Lucey}, J.~R., {Barris}, B.~J., {Hudson}, M.~J., \&
  {Tonry}, J.~L. 2001, \mnras, 327, 1004,
  \dodoi{10.1046/j.1365-8711.2001.04800.x}

\bibitem[{{Blandford} \& {McKee}(1982)}]{blanford1982}
{Blandford}, R.~D., \& {McKee}, C.~F. 1982, \apj, 255, 419,
  \dodoi{10.1086/159843}

\bibitem[{{Cackett} {et~al.}(2021){Cackett}, {Bentz}, \& {Kara}}]{cackett2021}
{Cackett}, E.~M., {Bentz}, M.~C., \& {Kara}, E. 2021, iScience, 24, 102557,
  \dodoi{10.1016/j.isci.2021.102557}

\bibitem[{{Cappellari}(2002)}]{cappellari2002}
{Cappellari}, M. 2002, \mnras, 333, 400,
  \dodoi{10.1046/j.1365-8711.2002.05412.x}

\bibitem[{{Cappellari}(2008)}]{cappellari2008}
---. 2008, \mnras, 390, 71, \dodoi{10.1111/j.1365-2966.2008.13754.x}

\bibitem[{{Cappellari}(2017)}]{cappellari2017}
---. 2017, \mnras, 466, 798, \dodoi{10.1093/mnras/stw3020}

\bibitem[{{Cappellari} \& {Copin}(2003)}]{cappellari2003}
{Cappellari}, M., \& {Copin}, Y. 2003, \mnras, 342, 345,
  \dodoi{10.1046/j.1365-8711.2003.06541.x}

\bibitem[{{Cappellari} \& {Emsellem}(2004)}]{cappellari2004}
{Cappellari}, M., \& {Emsellem}, E. 2004, \pasp, 116, 138,
  \dodoi{10.1086/381875}

\bibitem[{{Cappellari} {et~al.}(2011){Cappellari}, {Emsellem}, {Krajnovi{\'c}},
  {McDermid}, {Scott}, {Verdoes Kleijn}, {Young}, {Alatalo}, {Bacon}, {Blitz},
  {Bois}, {Bournaud}, {Bureau}, {Davies}, {Davis}, {de Zeeuw}, {Duc},
  {Khochfar}, {Kuntschner}, {Lablanche}, {Morganti}, {Naab}, {Oosterloo},
  {Sarzi}, {Serra}, \& {Weijmans}}]{cappellari2011}
{Cappellari}, M., {Emsellem}, E., {Krajnovi{\'c}}, D., {et~al.} 2011, \mnras,
  413, 813, \dodoi{10.1111/j.1365-2966.2010.18174.x}

\bibitem[{{Cappellari} {et~al.}(2013{\natexlab{a}}){Cappellari}, {McDermid},
  {Alatalo}, {Blitz}, {Bois}, {Bournaud}, {Bureau}, {Crocker}, {Davies},
  {Davis}, {de Zeeuw}, {Duc}, {Emsellem}, {Khochfar}, {Krajnovi{\'c}},
  {Kuntschner}, {Morganti}, {Naab}, {Oosterloo}, {Sarzi}, {Scott}, {Serra},
  {Weijmans}, \& {Young}}]{cappellari2013b}
{Cappellari}, M., {McDermid}, R.~M., {Alatalo}, K., {et~al.}
  2013{\natexlab{a}}, \mnras, 432, 1862, \dodoi{10.1093/mnras/stt644}

\bibitem[{{Cappellari} {et~al.}(2013{\natexlab{b}}){Cappellari}, {Scott},
  {Alatalo}, {Blitz}, {Bois}, {Bournaud}, {Bureau}, {Crocker}, {Davies},
  {Davis}, {de Zeeuw}, {Duc}, {Emsellem}, {Khochfar}, {Krajnovi{\'c}},
  {Kuntschner}, {McDermid}, {Morganti}, {Naab}, {Oosterloo}, {Sarzi}, {Serra},
  {Weijmans}, \& {Young}}]{cappellari2013a}
{Cappellari}, M., {Scott}, N., {Alatalo}, K., {et~al.} 2013{\natexlab{b}},
  \mnras, 432, 1709, \dodoi{10.1093/mnras/stt562}

\bibitem[{{Clements}(1983)}]{clements1983}
{Clements}, E.~D. 1983, \mnras, 204, 811, \dodoi{10.1093/mnras/204.3.811}

\bibitem[{{Contopoulos}(1956)}]{contopoulos1956}
{Contopoulos}, G. 1956, \zap, 39, 126

\bibitem[{{Davies} {et~al.}(2006){Davies}, {Thomas}, {Genzel}, {M{\"u}ller
  S{\'a}nchez}, {Tacconi}, {Sternberg}, {Eisenhauer}, {Abuter}, {Saglia}, \&
  {Bender}}]{davies2006}
{Davies}, R.~I., {Thomas}, J., {Genzel}, R., {et~al.} 2006, \apj, 646, 754,
  \dodoi{10.1086/504963}

\bibitem[{{Davis} {et~al.}(2013){Davis}, {Bureau}, {Cappellari}, {Sarzi}, \&
  {Blitz}}]{davis2013}
{Davis}, T.~A., {Bureau}, M., {Cappellari}, M., {Sarzi}, M., \& {Blitz}, L.
  2013, \nat, 494, 328, \dodoi{10.1038/nature11819}

\bibitem[{{De Rosa} {et~al.}(2018){De Rosa}, {Fausnaugh}, {Grier}, {Peterson},
  {Denney}, {Horne}, {Bentz}, {Ciroi}, {Dalla Bont{\`a}}, {Joner}, {Kaspi},
  {Kochanek}, {Pogge}, {Sergeev}, {Vestergaard}, {Adams}, {Antognini}, {Araya
  Salvo}, {Armstrong}, {Bae}, {Barth}, {Beatty}, {Bhattacharjee}, {Borman},
  {Boroson}, {Bottorff}, {Brown}, {Brown}, {Brotherton}, {Coker}, {Clanton},
  {Cracco}, {Crawford}, {Croxall}, {Eftekharzadeh}, {Eracleous}, {Fiorenza},
  {Frassati}, {Hawkins}, {Henderson}, {Holoien}, {Hutchison}, {Kellar},
  {Kilerci-Eser}, {Kim}, {King}, {La Mura}, {Laney}, {Li}, {Lochhaas}, {Ma},
  {MacInnis}, {Manne-Nicholas}, {Mason}, {McGraw}, {Mogren}, {Montouri},
  {Moody}, {Mosquera}, {Mudd}, {Musso}, {Nazarov}, {Nguyen}, {Ochner},
  {Okhmat}, {Onken}, {Ou-Yang}, {Pancoast}, {Pei}, {Penny}, {Poleski},
  {Portaluri}, {Prieto}, {Price-Whelan}, {Pulatova}, {Rafter}, {Roettenbacher},
  {Romero-Colmenero}, {Runnoe}, {Schimoia}, {Shappee}, {Sherf}, {Simonian},
  {Siviero}, {Skowron}, {Skowron}, {Somers}, {Spencer}, {Starkey}, {Stevens},
  {Stoll}, {Tamajo}, {Tayar}, {van Saders}, {Valenti}, {Villanueva},
  {Villforth}, {Weiss}, {Winkler}, {Zastrow}, {Zhu}, \& {Zu}}]{derosa2018}
{De Rosa}, G., {Fausnaugh}, M.~M., {Grier}, C.~J., {et~al.} 2018, \apj, 866,
  133, \dodoi{10.3847/1538-4357/aadd11}

\bibitem[{{de Vaucouleurs} {et~al.}(1991){de Vaucouleurs}, {de Vaucouleurs},
  {Corwin}, {Buta}, {Paturel}, \& {Fouque}}]{devaucouleurs1991}
{de Vaucouleurs}, G., {de Vaucouleurs}, A., {Corwin}, Herold~G., J., {et~al.}
  1991, {Third Reference Catalogue of Bright Galaxies}

\bibitem[{{Denney} {et~al.}(2010){Denney}, {Peterson}, {Pogge}, {Adair},
  {Atlee}, {Au-Yong}, {Bentz}, {Bird}, {Brokofsky}, {Chisholm}, {Comins},
  {Dietrich}, {Doroshenko}, {Eastman}, {Efimov}, {Ewald}, {Ferbey}, {Gaskell},
  {Hedrick}, {Jackson}, {Klimanov}, {Klimek}, {Kruse}, {Lad{\'e}route}, {Lamb},
  {Leighly}, {Minezaki}, {Nazarov}, {Onken}, {Petersen}, {Peterson},
  {Poindexter}, {Sakata}, {Schlesinger}, {Sergeev}, {Skolski}, {Stieglitz},
  {Tobin}, {Unterborn}, {Vestergaard}, {Watkins}, {Watson}, \&
  {Yoshii}}]{denney2010}
{Denney}, K.~D., {Peterson}, B.~M., {Pogge}, R.~W., {et~al.} 2010, \apj, 721,
  715, \dodoi{10.1088/0004-637X/721/1/715}

\bibitem[{{Emsellem} {et~al.}(2004){Emsellem}, {Cappellari}, {Peletier},
  {McDermid}, {Bacon}, {Bureau}, {Copin}, {Davies}, {Krajnovi{\'c}},
  {Kuntschner}, {Miller}, \& {de Zeeuw}}]{emsellem2004}
{Emsellem}, E., {Cappellari}, M., {Peletier}, R.~F., {et~al.} 2004, \mnras,
  352, 721, \dodoi{10.1111/j.1365-2966.2004.07948.x}

\bibitem[{{Erwin} {et~al.}(2018){Erwin}, {Thomas}, {Saglia}, {Fabricius},
  {Rusli}, {Seitz}, \& {Bender}}]{erwin2018}
{Erwin}, P., {Thomas}, J., {Saglia}, R.~P., {et~al.} 2018, \mnras, 473, 2251,
  \dodoi{10.1093/mnras/stx2499}

\bibitem[{{Event Horizon Telescope Collaboration} {et~al.}(2019){Event Horizon
  Telescope Collaboration}, {Akiyama}, {Alberdi}, {Alef}, {Asada}, {Azulay},
  {Baczko}, {Ball}, {Balokovi{\'c}}, {Barrett}, {Bintley}, {Blackburn},
  {Boland}, {Bouman}, {Bower}, {Bremer}, {Brinkerink}, {Brissenden}, {Britzen},
  {Broderick}, {Broguiere}, {Bronzwaer}, {Byun}, {Carlstrom}, {Chael}, {Chan},
  {Chatterjee}, {Chatterjee}, {Chen}, {Chen}, {Cho}, {Christian}, {Conway},
  {Cordes}, {Crew}, {Cui}, {Davelaar}, {De Laurentis}, {Deane}, {Dempsey},
  {Desvignes}, {Dexter}, {Doeleman}, {Eatough}, {Falcke}, {Fish}, {Fomalont},
  {Fraga-Encinas}, {Friberg}, {Fromm}, {G{\'o}mez}, {Galison}, {Gammie},
  {Garc{\'\i}a}, {Gentaz}, {Georgiev}, {Goddi}, {Gold}, {Gu}, {Gurwell},
  {Hada}, {Hecht}, {Hesper}, {Ho}, {Ho}, {Honma}, {Huang}, {Huang}, {Hughes},
  {Ikeda}, {Inoue}, {Issaoun}, {James}, {Jannuzi}, {Janssen}, {Jeter}, {Jiang},
  {Johnson}, {Jorstad}, {Jung}, {Karami}, {Karuppusamy}, {Kawashima},
  {Keating}, {Kettenis}, {Kim}, {Kim}, {Kim}, {Kino}, {Koay}, {Koch}, {Koyama},
  {Kramer}, {Kramer}, {Krichbaum}, {Kuo}, {Lauer}, {Lee}, {Li}, {Li},
  {Lindqvist}, {Liu}, {Liuzzo}, {Lo}, {Lobanov}, {Loinard}, {Lonsdale}, {Lu},
  {MacDonald}, {Mao}, {Markoff}, {Marrone}, {Marscher}, {Mart{\'\i}-Vidal},
  {Matsushita}, {Matthews}, {Medeiros}, {Menten}, {Mizuno}, {Mizuno}, {Moran},
  {Moriyama}, {Moscibrodzka}, {M{\"u}ller}, {Nagai}, {Nagar}, {Nakamura},
  {Narayan}, {Narayanan}, {Natarajan}, {Neri}, {Ni}, {Noutsos}, {Okino},
  {Olivares}, {Oyama}, {{\"O}zel}, {Palumbo}, {Patel}, {Pen}, {Pesce},
  {Pi{\'e}tu}, {Plambeck}, {PopStefanija}, {Porth}, {Prather},
  {Preciado-L{\'o}pez}, {Psaltis}, {Pu}, {Ramakrishnan}, {Rao}, {Rawlings},
  {Raymond}, {Rezzolla}, {Ripperda}, {Roelofs}, {Rogers}, {Ros}, {Rose},
  {Roshanineshat}, {Rottmann}, {Roy}, {Ruszczyk}, {Ryan}, {Rygl},
  {S{\'a}nchez}, {S{\'a}nchez-Arguelles}, {Sasada}, {Savolainen}, {Schloerb},
  {Schuster}, {Shao}, {Shen}, {Small}, {Sohn}, {SooHoo}, {Tazaki}, {Tiede},
  {Tilanus}, {Titus}, {Toma}, {Torne}, {Trent}, {Trippe}, {Tsuda}, {van
  Bemmel}, {van Langevelde}, {van Rossum}, {Wagner}, {Wardle}, {Weintroub},
  {Wex}, {Wharton}, {Wielgus}, {Wong}, {Wu}, {Young}, {Young}, {Younsi},
  {Yuan}, {Yuan}, {Zensus}, {Zhao}, {Zhao}, {Zhu}, {Farah}, {Meyer-Zhao},
  {Michalik}, {Nadolski}, {Nishioka}, {Pradel}, {Primiani}, {Souccar},
  {Vertatschitsch}, \& {Yamaguchi}}]{eht2019}
{Event Horizon Telescope Collaboration}, {Akiyama}, K., {Alberdi}, A., {et~al.}
  2019, \apjl, 875, L6, \dodoi{10.3847/2041-8213/ab1141}

\bibitem[{{Falc{\'o}n-Barroso} {et~al.}(2011){Falc{\'o}n-Barroso},
  {S{\'a}nchez-Bl{\'a}zquez}, {Vazdekis}, {Ricciardelli}, {Cardiel}, {Cenarro},
  {Gorgas}, \& {Peletier}}]{falconbarroso2011}
{Falc{\'o}n-Barroso}, J., {S{\'a}nchez-Bl{\'a}zquez}, P., {Vazdekis}, A.,
  {et~al.} 2011, \aap, 532, A95, \dodoi{10.1051/0004-6361/201116842}

\bibitem[{{Ferrarese} \& {Merritt}(2000)}]{ferrarese2000}
{Ferrarese}, L., \& {Merritt}, D. 2000, \apjl, 539, L9, \dodoi{10.1086/312838}

\bibitem[{{Gallimore} {et~al.}(2006){Gallimore}, {Axon}, {O'Dea}, {Baum}, \&
  {Pedlar}}]{gallimore2006}
{Gallimore}, J.~F., {Axon}, D.~J., {O'Dea}, C.~P., {Baum}, S.~A., \& {Pedlar},
  A. 2006, \aj, 132, 546, \dodoi{10.1086/504593}

\bibitem[{{Garc{\'\i}a-Lorenzo} {et~al.}(2015){Garc{\'\i}a-Lorenzo},
  {M{\'a}rquez}, {Barrera-Ballesteros}, {Masegosa}, {Husemann},
  {Falc{\'o}n-Barroso}, {Lyubenova}, {S{\'a}nchez}, {Walcher}, {Mast},
  {Garc{\'\i}a-Benito}, {M{\'e}ndez-Abreu}, {van de Ven}, {Spekkens}, {Holmes},
  {Monreal-Ibero}, {del Olmo}, {Ziegler}, {Bland-Hawthorn},
  {S{\'a}nchez-Bl{\'a}zquez}, {Iglesias-P{\'a}ramo}, {Aguerri}, {Papaderos},
  {Gomes}, {Marino}, {Gonz{\'a}lez Delgado}, {Cortijo-Ferrero},
  {L{\'o}pez-S{\'a}nchez}, {Bekerait{\.{e}}}, {Wisotzki}, \&
  {Bomans}}]{garcia2015}
{Garc{\'\i}a-Lorenzo}, B., {M{\'a}rquez}, I., {Barrera-Ballesteros}, J.~K.,
  {et~al.} 2015, \aap, 573, A59, \dodoi{10.1051/0004-6361/201423485}

\bibitem[{{Gebhardt} {et~al.}(2011){Gebhardt}, {Adams}, {Richstone}, {Lauer},
  {Faber}, {G{\"u}ltekin}, {Murphy}, \& {Tremaine}}]{gebhardt2011}
{Gebhardt}, K., {Adams}, J., {Richstone}, D., {et~al.} 2011, \apj, 729, 119,
  \dodoi{10.1088/0004-637X/729/2/119}

\bibitem[{{Gebhardt} \& {Thomas}(2009)}]{gebhardt2009}
{Gebhardt}, K., \& {Thomas}, J. 2009, \apj, 700, 1690,
  \dodoi{10.1088/0004-637X/700/2/1690}

\bibitem[{{Gebhardt} {et~al.}(2000){Gebhardt}, {Bender}, {Bower}, {Dressler},
  {Faber}, {Filippenko}, {Green}, {Grillmair}, {Ho}, {Kormendy}, {Lauer},
  {Magorrian}, {Pinkney}, {Richstone}, \& {Tremaine}}]{gebhardt2000}
{Gebhardt}, K., {Bender}, R., {Bower}, G., {et~al.} 2000, \apjl, 539, L13,
  \dodoi{10.1086/312840}

\bibitem[{{Gebhardt} {et~al.}(2003){Gebhardt}, {Richstone}, {Tremaine},
  {Lauer}, {Bender}, {Bower}, {Dressler}, {Faber}, {Filippenko}, {Green},
  {Grillmair}, {Ho}, {Kormendy}, {Magorrian}, \& {Pinkney}}]{gebhardt2003}
{Gebhardt}, K., {Richstone}, D., {Tremaine}, S., {et~al.} 2003, \apj, 583, 92,
  \dodoi{10.1086/345081}

\bibitem[{{Genzel} {et~al.}(2000){Genzel}, {Pichon}, {Eckart}, {Gerhard}, \&
  {Ott}}]{genzel00}
{Genzel}, R., {Pichon}, C., {Eckart}, A., {Gerhard}, O.~E., \& {Ott}, T. 2000,
  \mnras, 317, 348, \dodoi{10.1046/j.1365-8711.2000.03582.x}

\bibitem[{{Gerhard} \& {Binney}(1996)}]{gerhard1996}
{Gerhard}, O.~E., \& {Binney}, J.~J. 1996, \mnras, 279, 993,
  \dodoi{10.1093/mnras/279.3.993}

\bibitem[{{Ghez} {et~al.}(2000){Ghez}, {Morris}, {Becklin}, {Tanner}, \&
  {Kremenek}}]{ghez00}
{Ghez}, A.~M., {Morris}, M., {Becklin}, E.~E., {Tanner}, A., \& {Kremenek}, T.
  2000, \nat, 407, 349, \dodoi{10.1038/35030032}

\bibitem[{{GRAVITY Collaboration} {et~al.}(2022){GRAVITY Collaboration},
  {Abuter}, {Aimar}, {Amorim}, {Ball}, {Baub{\"o}ck}, {Berger}, {Bonnet},
  {Bourdarot}, {Brandner}, {Cardoso}, {Cl{\'e}net}, {Dallilar}, {Davies}, {de
  Zeeuw}, {Dexter}, {Drescher}, {Eisenhauer}, {F{\"o}rster Schreiber},
  {Foschi}, {Garcia}, {Gao}, {Gendron}, {Genzel}, {Gillessen}, {Habibi},
  {Haubois}, {Hei{\ss}el}, {Henning}, {Hippler}, {Horrobin}, {Jochum}, {Jocou},
  {Kaufer}, {Kervella}, {Lacour}, {Lapeyr{\`e}re}, {Le Bouquin}, {L{\'e}na},
  {Lutz}, {Ott}, {Paumard}, {Perraut}, {Perrin}, {Pfuhl}, {Rabien},
  {Shangguan}, {Shimizu}, {Scheithauer}, {Stadler}, {Stephens}, {Straub},
  {Straubmeier}, {Sturm}, {Tacconi}, {Tristram}, {Vincent}, {von Fellenberg},
  {Widmann}, {Wieprecht}, {Wiezorrek}, {Woillez}, {Yazici}, \&
  {Young}}]{gravity2022}
{GRAVITY Collaboration}, {Abuter}, R., {Aimar}, N., {et~al.} 2022, \aap, 657,
  L12, \dodoi{10.1051/0004-6361/202142465}

\bibitem[{{Grier} {et~al.}(2013){Grier}, {Martini}, {Watson}, {Peterson},
  {Bentz}, {Dasyra}, {Dietrich}, {Ferrarese}, {Pogge}, \& {Zu}}]{grier2013}
{Grier}, C.~J., {Martini}, P., {Watson}, L.~C., {et~al.} 2013, \apj, 773, 90,
  \dodoi{10.1088/0004-637X/773/2/90}

\bibitem[{{G{\"u}ltekin} {et~al.}(2009){G{\"u}ltekin}, {Richstone}, {Gebhardt},
  {Lauer}, {Tremaine}, {Aller}, {Bender}, {Dressler}, {Faber}, {Filippenko},
  {Green}, {Ho}, {Kormendy}, {Magorrian}, {Pinkney}, \&
  {Siopis}}]{gultekin2009}
{G{\"u}ltekin}, K., {Richstone}, D.~O., {Gebhardt}, K., {et~al.} 2009, \apj,
  698, 198, \dodoi{10.1088/0004-637X/698/1/198}

\bibitem[{{Guti{\'e}rrez} {et~al.}(2011){Guti{\'e}rrez}, {Erwin}, {Aladro}, \&
  {Beckman}}]{gutierrez2011}
{Guti{\'e}rrez}, L., {Erwin}, P., {Aladro}, R., \& {Beckman}, J.~E. 2011, \aj,
  142, 145, \dodoi{10.1088/0004-6256/142/5/145}

\bibitem[{{H{\"a}ring} \& {Rix}(2004)}]{haring2004}
{H{\"a}ring}, N., \& {Rix}, H.-W. 2004, \apjl, 604, L89, \dodoi{10.1086/383567}

\bibitem[{{Heckman} \& {Best}(2014)}]{heckman2014}
{Heckman}, T.~M., \& {Best}, P.~N. 2014, \araa, 52, 589,
  \dodoi{10.1146/annurev-astro-081913-035722}

\bibitem[{{Hicks} \& {Malkan}(2008)}]{hicks2008}
{Hicks}, E. K.~S., \& {Malkan}, M.~A. 2008, \apjs, 174, 31,
  \dodoi{10.1086/521650}

\bibitem[{{Holmberg}(1958)}]{holmberg1958}
{Holmberg}, E. 1958, Meddelanden fran Lunds Astronomiska Observatorium Serie
  II, 136, 1

\bibitem[{{Hubble}(1926)}]{hubble1926}
{Hubble}, E.~P. 1926, \apj, 64, 321, \dodoi{10.1086/143018}

\bibitem[{{Jeter} {et~al.}(2019){Jeter}, {Broderick}, \&
  {McNamara}}]{jeter2019}
{Jeter}, B., {Broderick}, A.~E., \& {McNamara}, B.~R. 2019, \apj, 882, 82,
  \dodoi{10.3847/1538-4357/ab3221}

\bibitem[{{Kakkad} {et~al.}(2017){Kakkad}, {Mainieri}, {Brusa}, {Padovani},
  {Carniani}, {Feruglio}, {Sargent}, {Husemann}, {Bongiorno}, {Bonzini},
  {Piconcelli}, {Silverman}, \& {Rujopakarn}}]{kakkad2017}
{Kakkad}, D., {Mainieri}, V., {Brusa}, M., {et~al.} 2017, \mnras, 468, 4205,
  \dodoi{10.1093/mnras/stx726}

\bibitem[{{Kochanek} \& {Rybicki}(1996)}]{kochanek1996}
{Kochanek}, C.~S., \& {Rybicki}, G.~B. 1996, \mnras, 280, 1257,
  \dodoi{10.1093/mnras/280.4.1257}

\bibitem[{{Kormendy} \& {Ho}(2013)}]{kormendy2013}
{Kormendy}, J., \& {Ho}, L.~C. 2013, \araa, 51, 511,
  \dodoi{10.1146/annurev-astro-082708-101811}

\bibitem[{{Koz{\l}owski}(2017)}]{kozlowski2017}
{Koz{\l}owski}, S. 2017, \apjs, 228, 9, \dodoi{10.3847/1538-4365/228/1/9}

\bibitem[{{Krajnovi{\'c}} {et~al.}(2006){Krajnovi{\'c}}, {Cappellari}, {de
  Zeeuw}, \& {Copin}}]{krajnovic2006}
{Krajnovi{\'c}}, D., {Cappellari}, M., {de Zeeuw}, P.~T., \& {Copin}, Y. 2006,
  \mnras, 366, 787, \dodoi{10.1111/j.1365-2966.2005.09902.x}

\bibitem[{{Krajnovi{\'c}} {et~al.}(2011){Krajnovi{\'c}}, {Emsellem},
  {Cappellari}, {Alatalo}, {Blitz}, {Bois}, {Bournaud}, {Bureau}, {Davies},
  {Davis}, {de Zeeuw}, {Khochfar}, {Kuntschner}, {Lablanche}, {McDermid},
  {Morganti}, {Naab}, {Oosterloo}, {Sarzi}, {Scott}, {Serra}, {Weijmans}, \&
  {Young}}]{krajnovic2011}
{Krajnovi{\'c}}, D., {Emsellem}, E., {Cappellari}, M., {et~al.} 2011, \mnras,
  414, 2923, \dodoi{10.1111/j.1365-2966.2011.18560.x}

\bibitem[{{Krajnovi{\'c}} {et~al.}(2013){Krajnovi{\'c}}, {Karick}, {Davies},
  {Naab}, {Sarzi}, {Emsellem}, {Cappellari}, {Serra}, {de Zeeuw}, {Scott},
  {McDermid}, {Weijmans}, {Davis}, {Alatalo}, {Blitz}, {Bois}, {Bureau},
  {Bournaud}, {Crocker}, {Duc}, {Khochfar}, {Kuntschner}, {Morganti},
  {Oosterloo}, \& {Young}}]{krajnovic2013}
{Krajnovi{\'c}}, D., {Karick}, A.~M., {Davies}, R.~L., {et~al.} 2013, \mnras,
  433, 2812, \dodoi{10.1093/mnras/stt905}

\bibitem[{{Macchetto} {et~al.}(1997){Macchetto}, {Marconi}, {Axon}, {Capetti},
  {Sparks}, \& {Crane}}]{macchetto1997}
{Macchetto}, F., {Marconi}, A., {Axon}, D.~J., {et~al.} 1997, \apj, 489, 579,
  \dodoi{10.1086/304823}

\bibitem[{{Marconi} \& {Hunt}(2003)}]{marconi2003}
{Marconi}, A., \& {Hunt}, L.~K. 2003, \apjl, 589, L21, \dodoi{10.1086/375804}

\bibitem[{{McConnell} {et~al.}(2013){McConnell}, {Chen}, {Ma}, {Greene},
  {Lauer}, \& {Gebhardt}}]{mcconnell2013}
{McConnell}, N.~J., {Chen}, S.-F.~S., {Ma}, C.-P., {et~al.} 2013, \apjl, 768,
  L21, \dodoi{10.1088/2041-8205/768/1/L21}

\bibitem[{{McGregor} {et~al.}(2003){McGregor}, {Hart}, {Conroy}, {Pfitzner},
  {Bloxham}, {Jones}, {Downing}, {Dawson}, {Young}, {Jarnyk}, \& {Van
  Harmelen}}]{mcgregor2003}
{McGregor}, P.~J., {Hart}, J., {Conroy}, P.~G., {et~al.} 2003, in Society of
  Photo-Optical Instrumentation Engineers (SPIE) Conference Series, Vol. 4841,
  Instrument Design and Performance for Optical/Infrared Ground-based
  Telescopes, ed. M.~{Iye} \& A.~F.~M. {Moorwood}, 1581--1591,
  \dodoi{10.1117/12.459448}

\bibitem[{{M{\'e}ndez-Abreu} {et~al.}(2008){M{\'e}ndez-Abreu}, {Aguerri},
  {Corsini}, \& {Simonneau}}]{mendezabreu2008}
{M{\'e}ndez-Abreu}, J., {Aguerri}, J.~A.~L., {Corsini}, E.~M., \& {Simonneau},
  E. 2008, \aap, 478, 353, \dodoi{10.1051/0004-6361:20078089}

\bibitem[{{Merrell} {et~al.}(2020){Merrell}, {Bentz}, \& {Walsh}}]{merrell2020}
{Merrell}, K.~A., {Bentz}, M.~C., \& {Walsh}, J.~L. 2020, Research Notes of the
  American Astronomical Society, 4, 250, \dodoi{10.3847/2515-5172/abd637}

\bibitem[{{Miyoshi} {et~al.}(1995){Miyoshi}, {Moran}, {Herrnstein},
  {Greenhill}, {Nakai}, {Diamond}, \& {Inoue}}]{miyoshi1995}
{Miyoshi}, M., {Moran}, J., {Herrnstein}, J., {et~al.} 1995, \nat, 373, 127,
  \dodoi{10.1038/373127a0}

\bibitem[{{Moultaka} {et~al.}(2004){Moultaka}, {Ilovaisky}, {Prugniel}, \&
  {Soubiran}}]{moultaka2004}
{Moultaka}, J., {Ilovaisky}, S.~A., {Prugniel}, P., \& {Soubiran}, C. 2004,
  \pasp, 116, 693, \dodoi{10.1086/422177}

\bibitem[{{Nguyen} {et~al.}(2017){Nguyen}, {Seth}, {den Brok}, {Neumayer},
  {Cappellari}, {Barth}, {Caldwell}, {Williams}, \& {Binder}}]{nguyen2017}
{Nguyen}, D.~D., {Seth}, A.~C., {den Brok}, M., {et~al.} 2017, \apj, 836, 237,
  \dodoi{10.3847/1538-4357/aa5cb4}

\bibitem[{{Onken} {et~al.}(2014){Onken}, {Valluri}, {Brown}, {McGregor},
  {Peterson}, {Bentz}, {Ferrarese}, {Pogge}, {Vestergaard}, {Storchi-Bergmann},
  \& {Riffel}}]{onken2014}
{Onken}, C.~A., {Valluri}, M., {Brown}, J.~S., {et~al.} 2014, \apj, 791, 37,
  \dodoi{10.1088/0004-637X/791/1/37}

\bibitem[{{Pahari} {et~al.}(2017){Pahari}, {McHardy}, {Mallick}, {Dewangan}, \&
  {Misra}}]{pahari2017}
{Pahari}, M., {McHardy}, I.~M., {Mallick}, L., {Dewangan}, G.~C., \& {Misra},
  R. 2017, \mnras, 470, 3239, \dodoi{10.1093/mnras/stx1455}

\bibitem[{{Panessa} {et~al.}(2020){Panessa}, {Castangia}, {Malizia}, {Bassani},
  {Tarchi}, {Bazzano}, \& {Ubertini}}]{panessa2020}
{Panessa}, F., {Castangia}, P., {Malizia}, A., {et~al.} 2020, \aap, 641, A162,
  \dodoi{10.1051/0004-6361/201937407}

\bibitem[{{Peng} {et~al.}(2002){Peng}, {Ho}, {Impey}, \& {Rix}}]{peng2002}
{Peng}, C.~Y., {Ho}, L.~C., {Impey}, C.~D., \& {Rix}, H.-W. 2002, \aj, 124,
  266, \dodoi{10.1086/340952}

\bibitem[{{Peng} {et~al.}(2010){Peng}, {Ho}, {Impey}, \& {Rix}}]{peng2010}
---. 2010, \aj, 139, 2097, \dodoi{10.1088/0004-6256/139/6/2097}

\bibitem[{{Peterson}(1993)}]{peterson1993}
{Peterson}, B.~M. 1993, \pasp, 105, 247, \dodoi{10.1086/133140}

\bibitem[{{Peterson} \& {Wandel}(1999)}]{peterson1999}
{Peterson}, B.~M., \& {Wandel}, A. 1999, \apjl, 521, L95,
  \dodoi{10.1086/312190}

\bibitem[{{Peterson} \& {Wandel}(2000)}]{peterson2000}
---. 2000, \apjl, 540, L13, \dodoi{10.1086/312862}

\bibitem[{{Prugniel} \& {Heraudeau}(1998)}]{prugniel1998}
{Prugniel}, P., \& {Heraudeau}, P. 1998, \aaps, 128, 299,
  \dodoi{10.1051/aas:1998142}

\bibitem[{{Roberts} {et~al.}(2021){Roberts}, {Bentz}, {Vasiliev}, {Valluri}, \&
  {Onken}}]{roberts2021}
{Roberts}, C.~A., {Bentz}, M.~C., {Vasiliev}, E., {Valluri}, M., \& {Onken},
  C.~A. 2021, \apj, 916, 25, \dodoi{10.3847/1538-4357/ac05b6}

\bibitem[{{S{\'a}nchez-Bl{\'a}zquez} {et~al.}(2006){S{\'a}nchez-Bl{\'a}zquez},
  {Peletier}, {Jim{\'e}nez-Vicente}, {Cardiel}, {Cenarro},
  {Falc{\'o}n-Barroso}, {Gorgas}, {Selam}, \& {Vazdekis}}]{sanchezblazquez2006}
{S{\'a}nchez-Bl{\'a}zquez}, P., {Peletier}, R.~F., {Jim{\'e}nez-Vicente}, J.,
  {et~al.} 2006, \mnras, 371, 703, \dodoi{10.1111/j.1365-2966.2006.10699.x}

\bibitem[{{Sandage} {et~al.}(1970){Sandage}, {Freeman}, \&
  {Stokes}}]{sandage1970}
{Sandage}, A., {Freeman}, K.~C., \& {Stokes}, N.~R. 1970, \apj, 160, 831,
  \dodoi{10.1086/150475}

\bibitem[{{Schlafly} \& {Finkbeiner}(2011)}]{schlafly2011}
{Schlafly}, E.~F., \& {Finkbeiner}, D.~P. 2011, \apj, 737, 103,
  \dodoi{10.1088/0004-637X/737/2/103}

\bibitem[{{Schlegel} {et~al.}(1998){Schlegel}, {Finkbeiner}, \&
  {Davis}}]{schlegel1998}
{Schlegel}, D.~J., {Finkbeiner}, D.~P., \& {Davis}, M. 1998, \apj, 500, 525,
  \dodoi{10.1086/305772}

\bibitem[{{Schmitt} \& {Kinney}(2000)}]{schmitt2000}
{Schmitt}, H.~R., \& {Kinney}, A.~L. 2000, \apjs, 128, 479,
  \dodoi{10.1086/313397}

\bibitem[{{Sch{\"o}del} {et~al.}(2002){Sch{\"o}del}, {Ott}, {Genzel},
  {et~al.}}]{schodel02}
{Sch{\"o}del}, R., {Ott}, T., {Genzel}, R., {et~al.} 2002, \nat, 419, 694,
  \dodoi{10.1038/nature01121}

\bibitem[{{Silk} \& {Rees}(1998)}]{silk1998}
{Silk}, J., \& {Rees}, M.~J. 1998, \aap, 331, L1.
\newblock \doarXiv{astro-ph/9801013}

\bibitem[{{Tonry} {et~al.}(2001){Tonry}, {Dressler}, {Blakeslee}, {Ajhar},
  {Fletcher}, {Luppino}, {Metzger}, \& {Moore}}]{tonry2001}
{Tonry}, J.~L., {Dressler}, A., {Blakeslee}, J.~P., {et~al.} 2001, \apj, 546,
  681, \dodoi{10.1086/318301}

\bibitem[{{Trippe} {et~al.}(2010){Trippe}, {Crenshaw}, {Deo}, {Dietrich},
  {Kraemer}, {Rafter}, \& {Turner}}]{trippe2010}
{Trippe}, M.~L., {Crenshaw}, D.~M., {Deo}, R.~P., {et~al.} 2010, \apj, 725,
  1749, \dodoi{10.1088/0004-637X/725/2/1749}

\bibitem[{{Tully} {et~al.}(2016){Tully}, {Courtois}, \& {Sorce}}]{tully2016}
{Tully}, R.~B., {Courtois}, H.~M., \& {Sorce}, J.~G. 2016, \aj, 152, 50,
  \dodoi{10.3847/0004-6256/152/2/50}

\bibitem[{{Tully} \& {Pierce}(2000)}]{tully2000}
{Tully}, R.~B., \& {Pierce}, M.~J. 2000, \apj, 533, 744, \dodoi{10.1086/308700}

\bibitem[{{Ulvestad} \& {Wilson}(1984)}]{ulvestad1984}
{Ulvestad}, J.~S., \& {Wilson}, A.~S. 1984, \apj, 285, 439,
  \dodoi{10.1086/162520}

\bibitem[{{Vacca} {et~al.}(2003){Vacca}, {Cushing}, \& {Rayner}}]{vacca2003}
{Vacca}, W.~D., {Cushing}, M.~C., \& {Rayner}, J.~T. 2003, \pasp, 115, 389,
  \dodoi{10.1086/346193}

\bibitem[{{Valdes} {et~al.}(2004){Valdes}, {Gupta}, {Rose}, {Singh}, \&
  {Bell}}]{valdes2004}
{Valdes}, F., {Gupta}, R., {Rose}, J.~A., {Singh}, H.~P., \& {Bell}, D.~J.
  2004, \apjs, 152, 251, \dodoi{10.1086/386343}

\bibitem[{{Valluri} {et~al.}(2004){Valluri}, {Merritt}, \&
  {Emsellem}}]{valluri2004}
{Valluri}, M., {Merritt}, D., \& {Emsellem}, E. 2004, \apj, 602, 66,
  \dodoi{10.1086/380896}

\bibitem[{{van den Bosch}(2016)}]{vandenbosch2016}
{van den Bosch}, R. C.~E. 2016, \apj, 831, 134,
  \dodoi{10.3847/0004-637X/831/2/134}

\bibitem[{{van den Bosch} \& {de Zeeuw}(2010)}]{vandenbosch2010}
{van den Bosch}, R. C.~E., \& {de Zeeuw}, P.~T. 2010, \mnras, 401, 1770,
  \dodoi{10.1111/j.1365-2966.2009.15832.x}

\bibitem[{{van der Marel} {et~al.}(1998){van der Marel}, {Cretton}, {de Zeeuw},
  \& {Rix}}]{vandermarel1998}
{van der Marel}, R.~P., {Cretton}, N., {de Zeeuw}, P.~T., \& {Rix}, H.-W. 1998,
  \apj, 493, 613, \dodoi{10.1086/305147}

\bibitem[{{Vasiliev}(2019)}]{vasiliev2019}
{Vasiliev}, E. 2019, \mnras, 482, 1525, \dodoi{10.1093/mnras/sty2672}

\bibitem[{{Vasiliev} \& {Valluri}(2020)}]{vasiliev2020}
{Vasiliev}, E., \& {Valluri}, M. 2020, \apj, 889, 39,
  \dodoi{10.3847/1538-4357/ab5fe0}

\bibitem[{{Verdoes Kleijn} {et~al.}(2006){Verdoes Kleijn}, {van der Marel}, \&
  {Noel-Storr}}]{verdoes2006}
{Verdoes Kleijn}, G.~A., {van der Marel}, R.~P., \& {Noel-Storr}, J. 2006, \aj,
  131, 1961, \dodoi{10.1086/500973}

\bibitem[{{Vincentelli} {et~al.}(2020){Vincentelli}, {Mastroserio}, {McHardy},
  {Ingram}, \& {Pahari}}]{vincentelli2020}
{Vincentelli}, F.~M., {Mastroserio}, G., {McHardy}, I., {Ingram}, A., \&
  {Pahari}, M. 2020, \mnras, 492, 1135, \dodoi{10.1093/mnras/stz3511}

\bibitem[{{Walsh} {et~al.}(2013){Walsh}, {Barth}, {Ho}, \& {Sarzi}}]{walsh2013}
{Walsh}, J.~L., {Barth}, A.~J., {Ho}, L.~C., \& {Sarzi}, M. 2013, \apj, 770,
  86, \dodoi{10.1088/0004-637X/770/2/86}

\bibitem[{{Wang} {et~al.}(2021){Wang}, {Yang}, {Fan}, {Hennawi}, {Barth},
  {Banados}, {Bian}, {Boutsia}, {Connor}, {Davies}, {Decarli}, {Eilers},
  {Farina}, {Green}, {Jiang}, {Li}, {Mazzucchelli}, {Nanni}, {Schindler},
  {Venemans}, {Walter}, {Wu}, \& {Yue}}]{wang2021}
{Wang}, F., {Yang}, J., {Fan}, X., {et~al.} 2021, \apjl, 907, L1,
  \dodoi{10.3847/2041-8213/abd8c6}

\bibitem[{{Weijmans} {et~al.}(2014){Weijmans}, {de Zeeuw}, {Emsellem},
  {Krajnovi{\'c}}, {Lablanche}, {Alatalo}, {Blitz}, {Bois}, {Bournaud},
  {Bureau}, {Cappellari}, {Crocker}, {Davies}, {Davis}, {Duc}, {Khochfar},
  {Kuntschner}, {McDermid}, {Morganti}, {Naab}, {Oosterloo}, {Sarzi}, {Scott},
  {Serra}, {Verdoes Kleijn}, \& {Young}}]{weijmans2014}
{Weijmans}, A.-M., {de Zeeuw}, P.~T., {Emsellem}, E., {et~al.} 2014, \mnras,
  444, 3340, \dodoi{10.1093/mnras/stu1603}

\bibitem[{{Winge} {et~al.}(2009){Winge}, {Riffel}, \&
  {Storchi-Bergmann}}]{winge2009}
{Winge}, C., {Riffel}, R.~A., \& {Storchi-Bergmann}, T. 2009, \apjs, 185, 186,
  \dodoi{10.1088/0067-0049/185/1/186}

\bibitem[{{Yuan} {et~al.}(2020){Yuan}, {Fausnaugh}, {Hoffmann}, {Macri},
  {Peterson}, {Riess}, {Bentz}, {Brown}, {Bont{\`a}}, {Davies}, {Rosa},
  {Ferrarese}, {Grier}, {Hicks}, {Onken}, {Pogge}, {Storchi-Bergmann}, \&
  {Vestergaard}}]{yuan2020}
{Yuan}, W., {Fausnaugh}, M.~M., {Hoffmann}, S.~L., {et~al.} 2020, \apj, 902,
  26, \dodoi{10.3847/1538-4357/abb377}

\end{thebibliography}
\bibliographystyle{aasjournal}

\end{document}